\documentclass[longauth]{aa}  
\usepackage{graphicx}
\usepackage{txfonts}
\usepackage{natbib}
\usepackage{longtable}
\usepackage{supertabular}
\bibpunct{(}{)}{;}{a}{}{,}
\newcommand{\tn}[1]{\textnormal{#1}}

\newcommand{\csr}{\chi^2\tn{/d.o.f.}=}
\newcommand{\kref}[1]{{\S} \ref{#1}}
\begin{document}
   \title{Photometry and spectroscopy of GRB 060526: A detailed study of the afterglow and host galaxy of a z=3.2 gamma-ray burst \thanks{Based in part on observations obtained with the European Southern Observatory's Very Large Telescope under proposals 077.D-0661 (PI: Vreeswijk) and 177.A-0591 (PI: Hjorth), as well as observations obtained with the NASA/ESA Hubble Space Telescope under proposal 11734 (PI: Levan).}}

   \author{
C. C. Th\"one		\inst{	1,2	} \and
D. A. Kann		\inst{	3	} \and
G. J\'ohannesson	\inst{	4,5	} \and
J. H. Selj			\inst{	6	} \and
A. O. Jaunsen		\inst{	6	} \and
J. P. U. Fynbo		\inst{	1	} \and
C. W. Akerlof		\inst{ 7	}\and
K. S. Baliyan		\inst{	8	} \and
C. Bartolini		\inst{	9	} \and
I. F. Bikmaev		\inst{	10	} \and
J. S. Bloom		\inst{	11	} \and
R. A. Burenin		\inst{	12	} \and
B. E. Cobb		\inst{	11,13	} \and
S. Covino			\inst{	2	} \and
P. A. Curran		\inst{	14,15	} \and
H. Dahle			\inst{ 6	} \and
A. Ferrero			\inst{ 16	} \and
S. Foley			\inst{ 16,17 } \and
J. French			\inst{ 16	} \and
A. S. Fruchter \inst{ 18	} \and
S. Ganesh		\inst{	8	} \and
J. F. Graham \inst{ 18	} \and
G. Greco			\inst{	9	} \and
A. Guarnieri 		\inst{ 9	} \and
L. Hanlon			\inst{ 16	} \and
J. Hjorth			\inst{ 1	} \and
M. Ibrahimov		\inst{	19	} \and
G. L. Israel		\inst{	20	} \and
P. Jakobsson		\inst{	21	} \and
M. Jel\'inek			\inst{	22	} \and
B. L. Jensen		\inst{ 1	} \and
U. G. J\o rgensen	\inst{	23,24	} \and
I. M. Khamitov		\inst{	25	} \and
T. S. Koch		\inst{	26	} \and
A. J. Levan		\inst{ 27 }\and
D. Malesani		\inst{	1	} \and
N. Masetti			\inst{	28	} \and
S. Meehan		\inst{ 16 } \and
G. Melady			\inst{ 16	} \and
D. Nanni			\inst{	20	} \and
J. N\"ar\"anen		\inst{ 29	} \and
E. Pakstiene		\inst{	30	} \and
M. N. Pavlinsky 	\inst{ 12	} \and
D. A. Perley		\inst{ 11	} \and
A. Piccioni		\inst{	9	} \and
G. Pizzichini		\inst{	28	} \and
A. Pozanenko		\inst{	12	} \and
P. W. A. Roming		\inst{ 26,31	}\and
W. Rujopakarn		\inst{	32	} \and
V. Rumyantsev		\inst{	33	} \and
E. S. Rykoff   \inst{	34,35	} \and
D. Sharapov		\inst{	19	} \and
D. Starr			\inst{	10	} \and
R. A. Sunyaev		\inst{	12,36 } \and
H. Swan		\inst{ 7	}\and
N. R. Tanvir	\inst{ 37	} \and
F. Terra			\inst{ 38	} \and
P. M. Vreeswijk		\inst{ 1	}\and
A. C. Wilson		\inst{	39	}\and
S. A. Yost		\inst{	40	}\and
F. Yuan		\inst{	7	}
	}

   \offprints{christina.thoene@brera.inaf.it}

\institute{
Dark Cosmology Centre, Niels Bohr Institute, University of Copenhagen, Juliane Maries Vej 30, 2100 Copenhagen, Denmark\\
\email{christina.thoene@brera.inaf.it}
\and
INAF - Osservatorio Astronomico di Brera, via Bianchi 46, 23807 Merate (Lc), Italy
\and
Th\"uringer Landessternwarte Tautenburg, Sternwarte 5, 07778 Tautenburg, Germany
\and
Science Institute, University of Iceland, Dunhaga 3, 107 Reykjav\'ik, Iceland
\and
Hansen Experimental Physics Laboratory, Stanford University, Stanford, CA 94305
\and
Institute of Theoretical Astrophysics, University of Oslo, P.O. Box 1029 Blindern, N-0315 Oslo, Norway
\and
Department of Physics, University of Michigan, Ann Arbor, MI 48109, USA
\and
Physical Research Laboratory, Ahmedabad--380 009, India
\and
Dipartimento di Astronomia, Universita di Bologna, via Ranzani 1, 40127 Bologna, Italy
\and
Kazan Federal University and Academy of Sciences of Tatarstan, Kazan, Russia
\and
Department of Astronomy, 601 Campbell Hall, University of California, Berkeley, CA 94720--3411
\and
Space Research Institute (IKI), 84/32 Profsoyuznaya Str, Moscow 117997, Russia
\and
Department of Astronomy, Yale University, P.O. Box 208101, New Haven, CT 06520
\and
Astronomical Institute ``Anton Pannekoek'', University of Amsterdam, Kruislaan 403, 1098 SJ Amsterdam, The Netherlands
\and
Mullard Space Science Laboratory, University College London, Holmbury St. Mary, Dorking, Surrey RH5 6NT
\and
UCD School of Physics, University College Dublin, Belfield, Dublin 4, Ireland
\and
Max-Planck-Institute for Extraterrestrial Physics, Giessenbachstra\ss e, 85748 Garching, Germany
\and
Space Telescope Science Institute, 3700 San Martin Drive, Baltimore, MD21218 USA
\and
Ulugh Beg Astronomical Institute, 33 Astronomical Str., Tashkent 700052, Uzbekistan
\and
INAF-Osservatorio Astronomico di Roma, Via Frascati 33, 00040 Monteporzio Catone, Italy
\and
Centre for Astrophysics and Cosmology, Science Institute, University of Iceland, Dunhagi 5, 107 Reykjav\'ik, Iceland
\and
Instituto de Astrof\'isica de Andaluc\'ia (IAA-CSIC), P.O. Box 3.004, E-18.080 Granada, Spain
\and
Niels Bohr Institute, University of Copenhagen, Juliane Maries Vej 30, 2100 Copenhagen, Denmark
\and
Centre for Star and Planet Formation, Geological Museum, \O{}ster Voldgade 5-7, 1350 Copenhagen, Denmark.
\and
T\"UB\.ITAK National Observatory, Antalya, Turkey
\and
Department of Astronomy and Astrophysics, Pennsylvania State University, 525 Davey Lab, University Park, PA 16802
\and
Department of Physics, Univ. of Warwick, Coventry, CV4 7AL, UK
\and
INAF - Istituto di Astrofisica Spaziale e Fisica Cosmica di Bologna, Via Gobetti 101, 40129 Bologna, Italy
\and
University of Helsinki Observatory, P.O. Box 14 (T\"ahtitorninm\"aki), 00014, Finland
\and
VU Institute of Theoretical Physics and Astronomy, A. Gostauto 12, 01108 Vilnius, Lithuania
\and
Space Science Department, Southwest Research Institute, San Antonio, TX 78238, USA
\and
Steward Observatory, University of Arizona, 933 North Cherry Avenue, Tucson, AZ 85721, USA
\and
Crimean Astrophysical Observatory, Nauchny, Crimea, 98409, Ukraine
\and
Physics Department, University of California at Santa Barbara, 2233B Broida Hall, Santa Barbara, CA 93106, USA
\and
TABASGO Fellow
\and
Max-Plank-Institut f\"ur Astrophysik, Karl-Schwarzschild Stra\ss e 1, 85741 Garching, Germany
\and
Department of Physics and Astronomy, University of Leicester, University Road, Leicester, LE1 7RH, UK
\and
Second University of Roma ``Tor Vergata'', Italy
\and
Department of Astronomy, University of Texas at Austin, Austin TX, 78712
\and
Department of Physics, College of St. Benedict/St. John's University, Collegeville, MN 56321, USA
}

\date{Received 8 June 2008 / accepted 17 July 2010}

\begin{abstract}
{}
{With this paper we want to investigate the highly variable afterglow light curve and environment of gamma-ray burst (GRB) 060526 at $z=3.221$.}
{We present one of the largest photometric datasets ever obtained for a GRB afterglow, consisting of multi-color photometric data from the ultraviolet to the near infrared. The data set contains 412 data points in total to which we add additional data from the literature. Furthermore, we present low-resolution high signal-to-noise spectra of the afterglow. The afterglow light curve is modeled with both an analytical model using broken power law fits and with a broad-band numerical model which includes energy injections. The absorption lines detected in the spectra are used to derive column densities using a multi-ion single-component curve-of-growth analysis from which we derive the metallicity of the host of GRB 060526.}
{The temporal behaviour of the afterglow follows a double broken power law with breaks at $t=0.090\pm0.005$ and $t=2.401\pm0.061$ days. It shows deviations from the smooth set of power laws that can be modeled by additional energy injections from the central engine, although some significant microvariability remains. The broadband spectral-energy distribution of the afterglow shows no significant extinction along the line of sight. The metallicity derived from \ion{S}{II} and \ion{Fe}{II} of [S/H] = --0.57 $\pm$0.25 and [Fe/H] = --1.09$\pm$0.24 is relatively high for a galaxy at that redshift but comparable to the metallicity of other GRB hosts at similar redshifts. At the position of the afterglow, no host is detected to F775W(AB) = 28.5 mag with the HST, implying an absolute magnitude of the host M(1500 \AA{})$>$--18.3 mag which is fainter than most long-duration hosts, although the GRB may be associated with a faint galaxy at a distance of 11 kpc.}
{}
\end{abstract}


\authorrunning{C. C. Th\"one et al.}
\titlerunning{Photometry and Spectroscopy of GRB 060526}
\maketitle
%

\section{Introduction}

Gamma-ray bursts (GRBs) and their afterglows offer a powerful tool to probe the high-redshift universe, both through photometry and spectroscopy. The standard fireball model of GRB afterglows \citep[see][for recent reviews on the topic]{Zhang07, Meszaros06, Gehrels09} predicts a smooth temporal evolution, and the resulting afterglow light curve can be empirically described by a joint smoothly-broken power law, the so-called Beuermann equation \citep{Beuermann1999}. For many GRBs in the pre-{\it Swift} era, when  temporally dense afterglow photometry was obtained, the afterglow evolution was found to be smooth as, e.g., in GRB 020813 \citep{LaursenStanek2003}, GRB 030226 \citep{Klose2004} and GRB 041006 \citep{Stanek2005} \citep[for the complete pre-\emph{Swift} sample see][]{ZKK}. Out of the total sample of 59 afterglows of that paper, only four of the GRBs analysed showed significant deviations from the expected decay (though we note that about half of the 59 afterglows were not sampled well enough to allow any conclusions). One might be explained by microlensing \citep[GRB 000301C,][]{GLS2000} and another one by assuming an  inhomogeneously emitting surface \citep[GRB 011211,][]{JakobssonNA2004}. The other two GRBs which are the only ones showing long-lasting strong deviations are GRB 021004 \citep[e.g.][and references therein]{deUgarte2005} and GRB 030329 \citep[e.g.][and references therein]{Lipkin2004}, which incidentally also have the densest optical monitoring.

With the launch of the \emph{Swift} satellite and its rapid localization capabilities \citep{Gehrels2004}, the number of highly variable light curves has increased considerably, though there are still examples of very smooth light curves (e.g. that of GRB 080210, which showed a smooth behaviour with a high sampling rate of 1\,s, A. De Cia, priv. comm.). Some light curves show small, achromatic bumps overlying the smooth power law decay (e.g., GRB 050502A, \citealt{Guidorzi05}; GRB 061007, \citealt{Mundell07}; GRBs 090323 and 090328, \citealt{McBreenLAT}), early bumps with chromatic evolution (e.g., GRB 061126, \citealt{Perley08a}; GRB 071003, \citealt{Perley08b}), ``steps'' due to energy injection episodes (e.g., GRB 070125, \citealt{Updike08}; GRB 071010A, \citealt{Covino08}; GRB 080913, \citealt{Greiner080913}, GRB 090926A, \citealt{Rau090926A, CenkoLAT, Swenson090926A}) or powerful late-time rebrightenings of up to several magnitudes (e.g., GRB 050721, \citealt{Antonelli06}; GRB 060206, \citealt{Wozniak060206, Monfardini060206, Stanek060206}; GRB 070311, \citealt{Guidorzi07}; GRB 071003, \citealt{Perley08b}). The early time domain, which can now be routinely accessed by rapid follow-up in the \emph{Swift} era, has yielded more types of variability, like rising afterglows (e.g., GRB 060418, \citealt{Molinari07}; GRB 060605, \citealt{Ferrero08}; GRB 060607A, \citealt{Nysewander07}, \citealt{Molinari07}; GRB 081008, \citealt{Yuan10}; see \citealt{Oates09} and \citealt{Rykoff09} for further examples) and short-term variability directly linked to the prompt emission (e.g., GRB 041219A, \citealt{Vestrand041219A, Blake041219A}; GRB 050820A, \citealt{Vestrand050820A}; GRB 080319B, \citealt{Racusin08}; GRB 080129, \citealt{Greiner080129}). In all these cases, dense photometric follow-up during the periods of variability was needed to characterize the phenomena involved.

In addition their use in to studying the GRB phenomenon itself, GRB afterglows can be used to study their galactic environment through absorption line spectroscopy of material in the line-of-sight towards the GRB. Above a redshift of around $z\sim0.2$, resonant absorption lines from elements present in the interstellar medium (ISM) such as Mg, Zn, Fe, Si, C, and S are shifted into the optical regime and can be studied with ground-based telescopes. For 10 bursts between a redshift of $z=2$ to 6, the metallicity along the line-of-sight in the galaxy could be obtained so far \citep[e.g.][and references therein]{Savaglio06, Fynbo06a, Price07, Prochaska07, Ledoux09}. The values are usually below solar, but higher than for QSO absorbers at comparable redshifts, some of them even higher than theoretical limits for the formation of collapsars \citep{Woosley06}. The difference to QSO absorbers can be explained if GRB sightlines probe denser parts of the galaxy, or if GRBs reside in galaxies with higher masses and therefore higher metallicities \citep{Fynbo08, Pontzen10}. Both for QSO and GRB absorbers, there seems to be a metallicity evolution with redshift \citep{Savaglio06, Fynbo06a, Price07}, although the slope is different for the two samples. GRB hosts seem to show a low extinction along the line-of-sight \citep{Prochaska07}, however, relative abundances of heavier elements indicate that some of the ions must be depleted onto dust grains and the depletion pattern resembles the one found in the warm disc and halo of the Milky Way (MW; \citealt{Savaglio06}). How this can be explained together with the low extinction as also derived from the spectral energy distribution (SED) of the afterglow \citep{Kann2006, Starling07, Kann07}, is still an open question. One possible solution to this problem is destruction of the dust present in the line-of-sight by the GRB and afterglow radiation \citep[e.g.,][]{Waxman00, Perna02}, though no strong evidence has ever been found for this. Furthermore, there is clearly a strong observational bias involved, as those GRB afterglows with successful spectroscopy, especially in the case of high-resolution observations, are those which have only low extinctions and thus relatively bright afterglows \citep{Fynbo09, Kann07}, although rapid observations with large telescopes can achieve detailed spectroscopy of highly extinguished afterglows, as in the case of GRB 080607 \citep{Prochaska09, Sheffer09, Fynbo09}.

GRB 060526 was detected by the \emph{Swift} satellite on May 26.686458 (16:28:29.95 UT). The satellite slewed immediately to the burst, detecting both the X-ray and the optical afterglow \citep{CampanaGCN1}. The BAT instrument on \emph{Swift} measured two emission episodes. The first one lasted 13.8 s and consisted of two FRED (fast rise exponential decay) peaks, followed by a second symmetric peak between 230 and 270 s \citep{CampanaGCN2}. The second peak was coincident with a giant X-ray flare followed by a softer flare at 310 s \citep{CampanaGCN3} also detected in the $v$ band by the UVOT telescope on-board \emph{Swift} \citep{BrownGCN}. The gamma-ray fluence was $(4.9\pm0.6)\times10^{-7}$ erg cm$^{-2}$ during the first emission episode and $(5.9\pm0.6)\times10^{-7}$ erg cm$^{-2}$ during the second, the peak flux of the second episode was however only half of the peak flux of the first one. The photon index of the two epochs changed from $1.66\pm0.20$ to $2.07\pm0.18$, thus showing the typical hard-to-soft evolution \citep{MarkwardtGCN}. The \emph{Watcher} telescope provided the first ground based detection with $R\approx15$ mag \citep{FrenchGCN} 36.2 s after the trigger. \emph{ROTSE} observations showed a plateau for several thousand seconds after the GRB onset \citep{RykoffGCN}. A redshift of $z=3.21$ was determined by \cite{BergerGCN} with the Magellan/Clay telescope. The brightness of the optical afterglow allowed for a dense monitoring which revealed a complex light curve structure including several flares \citep{HalpernGCN1, HalpernGCN2} and a steepening attributed to a jet break \citep{ThoeneGCN}.

In this paper, we approach the analysis of GRB 060526 from two directions: Through modelling of the very detailed optical light curve and late optical imaging of the field to detect the host in Sec. 3, and analysis of low to medium resolution spectroscopic observations of absorption lines along the line of sight (Sec. 4). Throughout the paper, we follow the convention $F_\nu (t)\propto t^{-\alpha}\nu^{-\beta}$, and use WMAP concordant cosmology \citep{Spergel2003} with $H_0=71$km s$^{-1}$ Mpc$^{-1}$, $\Omega_M=0.27$, and $\Omega_{\Lambda}=0.73$. Uncertainties are given at 68\% confidence level for one parameter of interest unless stated otherwise.


\section{Observations}
\subsection{Photometry}

\onltab{1}{
\longtab{1}{
\begin{longtable}{llcccc}
\caption{\label{obslog} Broad band observations of GRB 060526, times given are the midpoint of the observations. Data have not been corrected for Galactic extinction. All data except the Keck/LRIS $g^\prime$ and the HST $F775W$ data points (in AB magnitudes) are in Vega magnitudes.}\\
\hline
\hline                
$\Delta$t [d] & exp [s] &instrument & filter & mag & error\\   
\hline        
\hline        
0.0250277	&	509.94	&	UVOT	&	$uvw2$	& $>	21.0	$ &	$\cdots$	\\
0.4692036	&	6180.78	&	UVOT	&	$uvw2$	& $>	22.7	$ &	$\cdots$	\\
1.8554459	&	2306.75	&	UVOT	&	$uvw2$	& $>	22.1	$ &	$\cdots$	\\
3.0522168	&	2212.79	&	UVOT	&	$uvw2$	& $>	22.1	$ &	$\cdots$	\\
3.9406911	&	1778.93	&	UVOT	&	$uvw2$	& $>	21.9	$ &	$\cdots$	\\
\hline									
0.0238351	&	471.09	&	UVOT	&	$uvm2$	& $>	20.6	$ &	$\cdots$	\\
0.8961627	&	3319.91	&	UVOT	&	$uvm2$	& $>	22.0	$ &	$\cdots$	\\
1.8577054	&	2306.9	&	UVOT	&	$uvm2$	& $>	21.8	$ &	$\cdots$	\\
3.0550826	&	2212.9	&	UVOT	&	$uvm2$	& $>	21.7	$ &	$\cdots$	\\
3.9434308	&	1775.78	&	UVOT	&	$uvm2$	& $>	21.6	$ &	$\cdots$	\\
\hline									
0.0246621	&	490.49	&	UVOT	&	$uvw1$	& $>	20.9	$ &	$\cdots$	\\
0.9543864	&	1030.99	&	UVOT	&	$uvw1$	& $>	21.4	$ &	$\cdots$	\\
1.8588450	&	2306.88	&	UVOT	&	$uvw1$	& $>	21.9	$ &	$\cdots$	\\
3.0565271	&	2192.57	&	UVOT	&	$uvw1$	& $>	21.9	$ &	$\cdots$	\\
3.9448082	&	1777.08	&	UVOT	&	$uvw1$	& $>	21.7	$ &	$\cdots$	\\
\hline									
0.0254789	&	490.45	&	UVOT	&	$u$	& $>	21.0	$ &	$\cdots$	\\
1.4941792	&	40.13	&	UVOT	&	$u$	& $>	19.3	$ &	$\cdots$	\\
1.8599737	&	2169.1	&	UVOT	&	$u$	& $>	21.9	$ &	$\cdots$	\\
3.0579600	&	2210.83	&	UVOT	&	$u$	& $>	21.9	$ &	$\cdots$	\\
3.9461762	&	1773.99	&	UVOT	&	$u$	& $>	21.8	$ &	$\cdots$	\\
\hline									
0.0122205	&	58.12	&	UVOT	&	$b$	&	18.270	&	0.160	\\
0.0392111	&	216.07	&	UVOT	&	$b$	&	18.710	&	0.110	\\
0.09394	&	180	&	Maidanak	&	$B$	&	19.518	&	0.207	\\
0.09898	&	180	&	Maidanak	&	$B$	&	19.557	&	0.135	\\
0.10625	&	180	&	Maidanak	&	$B$	&	19.432	&	0.136	\\
0.10925	&	180	&	Maidanak	&	$B$	&	19.505	&	0.107	\\
0.11421	&	900	&	Maidanak	&	$B$	&	19.546	&	0.179	\\
0.23242	&	100	&	RTT150	&	$B$	&	20.133	&	0.069	\\
0.23813	&	100	&	RTT150	&	$B$	&	20.321	&	0.083	\\
0.24525	&	100	&	RTT150	&	$B$	&	20.316	&	0.081	\\
0.26704	&	300	&	RTT150	&	$B$	&	20.395	&	0.064	\\
0.28883	&	300	&	RTT150	&	$B$	&	20.636	&	0.074	\\
0.31108	&	300	&	RTT150	&	$B$	&	20.649	&	0.073	\\
0.33371	&	300	&	RTT150	&	$B$	&	20.839	&	0.085	\\
1.10542	&	900	&	RTT150	&	$B$	&	22.072	&	0.097	\\
1.16521	&	900	&	Maidanak	&	$B$	&	22.013	&	0.079	\\
1.30279	&	900	&	RTT150	&	$B$	&	22.246	&	0.154	\\
1.4946505	&	30.99	&	UVOT	&	$b$	& $>	19.5	$ &	$\cdots$	\\
1.8609920	&	1909.33	&	UVOT	&	$b$	& $>	22.1	$ &	$\cdots$	\\
2.14626	&	2880	&	Maidanak	&	$B$	&	23.290	&	0.072	\\
2.35104	&	900	&	MOSCA	&	$B$	&	23.104	&	0.082	\\
3.0592114	&	1810.67	&	UVOT	&	$b$	& $>	22.1	$ &	$\cdots$	\\
3.13135	&	2700	&	Maidanak	&	$B$	&	24.198	&	0.115	\\
3.9474369	&	1536.37	&	UVOT	&	$b$	& $>	22.0	$ &	$\cdots$	\\
\hline									
3.73439	&	660	&	Keck/LRIS	&	$g^\prime$	&	23.952	&	0.022	\\
4.87114	&	660	&	Keck/LRIS	&	$g^\prime$	&	24.480	&	0.120	\\
\hline		
0.0021561	&	10	&	UVOT	&	$v$	&	16.610	&	0.280	\\
0.0022719	&	10	&	UVOT	&	$v$	&	16.540	&	0.190	\\
0.0023876	&	10	&	UVOT	&	$v$	&	16.480	&	0.180	\\
0.0025034	&	10	&	UVOT	&	$v$	&	16.660	&	0.200	\\
0.0026191	&	10	&	UVOT	&	$v$	&	16.830	&	0.210	\\
0.0027348	&	10	&	UVOT	&	$v$	&	17.000	&	0.230	\\
0.0028506	&	10	&	UVOT	&	$v$	&	16.710	&	0.200	\\
0.0029663	&	10	&	UVOT	&	$v$	&	15.940	&	0.140	\\
0.0030821	&	10	&	UVOT	&	$v$	&	16.350	&	0.170	\\
0.0031978	&	10	&	UVOT	&	$v$	&	16.610	&	0.190	\\
0.0033135	&	10	&	UVOT	&	$v$	&	16.940	&	0.220	\\
0.0034293	&	10	&	UVOT	&	$v$	&	16.670	&	0.200	\\
0.0035450	&	10	&	UVOT	&	$v$	&	17.200	&	0.260	\\
0.0036608	&	10	&	UVOT	&	$v$	&	17.140	&	0.250	\\
0.0037765	&	10	&	UVOT	&	$v$	&	17.570	&	0.300	\\
0.0038923	&	10	&	UVOT	&	$v$	&	18.000	&	0.380	\\
0.0040080	&	10	&	UVOT	&	$v$	&	17.340	&	0.270	\\
0.0041237	&	10	&	UVOT	&	$v$	&	17.010	&	0.230	\\
0.0042395	&	10	&	UVOT	&	$v$	&	17.270	&	0.260	\\
0.0043552	&	10	&	UVOT	&	$v$	&	16.920	&	0.220	\\
0.0044710	&	10	&	UVOT	&	$v$	&	17.520	&	0.300	\\
0.0045867	&	10	&	UVOT	&	$v$	&	17.580	&	0.310	\\
0.0049254	&	50	&	UVOT	&	$v$	&	17.780	&	0.150	\\
0.0054498	&	40	&	UVOT	&	$v$	&	17.720	&	0.160	\\
0.0059132	&	40	&	UVOT	&	$v$	&	17.710	&	0.160	\\
0.0064320	&	50	&	UVOT	&	$v$	&	17.900	&	0.160	\\
0.0082840	&	41.68	&	UVOT	&	$v$	&	17.770	&	0.160	\\
0.0105975	&	70	&	UVOT	&	$v$	&	18.160	&	0.150	\\
0.0112959	&	50	&	UVOT	&	$v$	&	17.900	&	0.160	\\
0.0118748	&	50	&	UVOT	&	$v$	&	17.930	&	0.160	\\
0.0125663	&	70	&	UVOT	&	$v$	&	18.240	&	0.160	\\
0.0136547	&	120	&	UVOT	&	$v$	&	18.820	&	0.160	\\
0.0169180	&	66.88	&	UVOT	&	$v$	&	18.250	&	0.160	\\
0.0356456	&	210.99	&	UVOT	&	$v$	&	18.370	&	0.120	\\
0.1052804	&	311.39	&	UVOT	&	$v$	&	18.820	&	0.140	\\
0.19284	&	60	&	DOLORES	&	$V$	&	19.235	&	0.087	\\
0.19443	&	60	&	DOLORES	&	$V$	&	19.338	&	0.065	\\
0.22210	&	120	&	MIRO	&	$V$	&	19.274	&	0.167	\\
0.23067	&	100	&	RTT150	&	$V$	&	19.302	&	0.046	\\
0.23583	&	100	&	RTT150	&	$V$	&	19.527	&	0.047	\\
0.24254	&	100	&	RTT150	&	$V$	&	19.364	&	0.045	\\
0.24818	&	1200	&	BFOSC	&	$V$	&	19.403	&	0.081	\\
0.26313	&	300	&	RTT150	&	$V$	&	19.584	&	0.040	\\
0.28487	&	300	&	RTT150	&	$V$	&	19.669	&	0.040	\\
0.28809	&	1200	&	BFOSC	&	$V$	&	19.655	&	0.078	\\
0.30667	&	300	&	RTT150	&	$V$	&	19.821	&	0.040	\\
0.32942	&	300	&	RTT150	&	$V$	&	19.861	&	0.048	\\
0.33005	&	1200	&	BFOSC	&	$V$	&	19.870	&	0.083	\\
0.41910	&	600	&	DFOSC	&	$V$	&	20.155	&	0.073	\\
0.43507	&	600	&	DFOSC	&	$V$	&	20.200	&	0.057	\\
1.10946	&	900	&	RTT150	&	$V$	&	21.027	&	0.055	\\
1.31288	&	900	&	RTT150	&	$V$	&	21.105	&	0.060	\\
1.40937	&	600	&	DFOSC	&	$V$	&	21.267	&	0.037	\\
1.41685	&	600	&	DFOSC	&	$V$	&	21.273	&	0.040	\\
1.42464	&	600	&	DFOSC	&	$V$	&	21.276	&	0.038	\\
1.42725	&	11382.11	&	UVOT	&	$v$	&	21.340	&	0.170	\\
2.32642	&	900	&	RTT150	&	$V$	&	22.203	&	0.123	\\
2.36561	&	600	&	DFOSC	&	$V$	&	22.279	&	0.062	\\
3.14483	&	3000	&	RTT150	&	$V$	&	22.818	&	0.094	\\
3.23404	&	3000	&	RTT150	&	$V$	&	22.834	&	0.094	\\
3.32054	&	3000	&	RTT150	&	$V$	&	23.123	&	0.154	\\
4.24058	&	7200	&	RTT150	&	$V$	&	23.508	&	0.121	\\
5.35387	&	8400	&	RTT150	&	$V$	&	24.322	&	0.211	\\
\hline									
0.0009974	&	10	&	UVOT	&	$white$	&	16.870	&	0.120	\\
0.0011133	&	10	&	UVOT	&	$white$	&	16.800	&	0.110	\\
0.0012292	&	10	&	UVOT	&	$white$	&	16.920	&	0.120	\\
0.0013451	&	10	&	UVOT	&	$white$	&	17.050	&	0.130	\\
0.0014609	&	10	&	UVOT	&	$white$	&	17.150	&	0.140	\\
0.0015767	&	10	&	UVOT	&	$white$	&	17.050	&	0.130	\\
0.0016925	&	10	&	UVOT	&	$white$	&	17.180	&	0.140	\\
0.0018083	&	10	&	UVOT	&	$white$	&	17.060	&	0.130	\\
0.0019241	&	10	&	UVOT	&	$white$	&	17.000	&	0.130	\\
0.0020399	&	9.76	&	UVOT	&	$white$	&	16.930	&	0.120	\\
0.0078848	&	9.77	&	UVOT	&	$white$	&	17.540	&	0.170	\\
0.0122853	&	107.98	&	UVOT	&	$white$	&	18.090	&	0.080	\\
0.0193600	&	29.28	&	UVOT	&	$white$	&	18.670	&	0.170	\\
0.0744556	&	196.63	&	UVOT	&	$white$	&	18.730	&	0.090	\\
\hline									
0.0005741	&	10	&	Watcher	&	$R_C$	&	16.110	&	0.430	\\
0.0013513	&	10	&	Watcher	&	$R_C$	& $>	16.080	$ &	$\cdots$	\\
0.0017219	&	10	&	Watcher	&	$R_C$	& $>	15.830	$ &	$\cdots$	\\
0.0025092	&	10	&	Watcher	&	$R_C$	& $>	16.240	$ &	$\cdots$	\\
0.0028681	&	10	&	Watcher	&	$R_C$	& $>	16.060	$ &	$\cdots$	\\
0.0036553	&	10	&	Watcher	&	$R_C$	& $>	16.050	$ &	$\cdots$	\\
0.0068646	&	60	&	Watcher	&	$R_C$	&	16.920	&	0.400	\\
0.0110924	&	60	&	Watcher	&	$R_C$	&	16.990	&	0.380	\\
0.0195091	&	60	&	Watcher	&	$R_C$	&	17.290	&	0.390	\\
0.0237226	&	60	&	Watcher	&	$R_C$	&	17.540	&	0.520	\\
0.0342566	&	120	&	Watcher	&	$R_C$	&	17.740	&	0.120	\\
0.0497130	&	240	&	Watcher	&	$R_C$	&	17.847	&	0.140	\\
0.0651205	&	240	&	Watcher	&	$R_C$	&	18.030	&	0.140	\\
0.0805210	&	240	&	Watcher	&	$R_C$	&	18.097	&	0.150	\\
0.09030	&	180	&	Maidanak	&	$R_C$	&	18.039	&	0.062	\\
0.09539	&	180	&	Maidanak	&	$R_C$	&	18.042	&	0.064	\\
0.095935	&	240	&	Watcher	&	$R_C$	&	18.047	&	0.140	\\
0.10079	&	180	&	Maidanak	&	$R_C$	&	18.117	&	0.067	\\
0.1113202	&	240	&	Watcher	&	$R_C$	&	18.247	&	0.180	\\
0.1267018	&	240	&	Watcher	&	$R_C$	&	18.137	&	0.150	\\
0.1457669	&	360	&	Watcher	&	$R_C$	&	18.363	&	0.140	\\
0.18919	&	60	&	DOLORES	&	$R_C$	&	18.627	&	0.109	\\
0.19079	&	60	&	DOLORES	&	$R_C$	&	18.763	&	0.097	\\
0.19165	&	180	&	MIRO	&	$R_C$	&	18.616	&	0.066	\\
0.19390	&	180	&	MIRO	&	$R_C$	&	18.646	&	0.068	\\
0.19615	&	180	&	MIRO	&	$R_C$	&	18.546	&	0.094	\\
0.19838	&	180	&	MIRO	&	$R_C$	&	18.775	&	0.046	\\
0.20914	&	600	&	BFOSC	&	$R_C$	&	18.723	&	0.087	\\
0.21225	&	180	&	MIRO	&	$R_C$	&	18.851	&	0.143	\\
0.21449	&	180	&	MIRO	&	$R_C$	&	18.778	&	0.273	\\
0.21897	&	180	&	MIRO	&	$R_C$	&	18.756	&	0.232	\\
0.22144	&	900	&	BFOSC	&	$R_C$	&	18.779	&	0.083	\\
0.22679	&	150	&	RTT150	&	$R_C$	&	18.576	&	0.076	\\
0.22892	&	150	&	RTT150	&	$R_C$	&	18.892	&	0.061	\\
0.23417	&	150	&	RTT150	&	$R_C$	&	18.831	&	0.060	\\
0.23428	&	900	&	BFOSC	&	$R_C$	&	18.821	&	0.078	\\
0.24058	&	150	&	RTT150	&	$R_C$	&	18.888	&	0.066	\\
0.24829	&	150	&	RTT150	&	$R_C$	&	18.881	&	0.071	\\
0.25054	&	150	&	RTT150	&	$R_C$	&	18.935	&	0.071	\\
0.25279	&	150	&	RTT150	&	$R_C$	&	18.960	&	0.070	\\
0.25504	&	150	&	RTT150	&	$R_C$	&	18.805	&	0.070	\\
0.25530	&	180	&	MIRO	&	$R_C$	&	18.815	&	0.155	\\
0.25729	&	150	&	RTT150	&	$R_C$	&	18.910	&	0.071	\\
0.25954	&	150	&	RTT150	&	$R_C$	&	18.917	&	0.073	\\
0.27050	&	150	&	RTT150	&	$R_C$	&	18.994	&	0.075	\\
0.27275	&	150	&	RTT150	&	$R_C$	&	19.105	&	0.081	\\
0.27468	&	900	&	BFOSC	&	$R_C$	&	19.017	&	0.074	\\
0.27500	&	150	&	RTT150	&	$R_C$	&	19.147	&	0.081	\\
0.27721	&	150	&	RTT150	&	$R_C$	&	18.993	&	0.074	\\
0.27946	&	150	&	RTT150	&	$R_C$	&	19.014	&	0.073	\\
0.28171	&	150	&	RTT150	&	$R_C$	&	18.986	&	0.075	\\
0.29208	&	150	&	RTT150	&	$R_C$	&	19.073	&	0.080	\\
0.29438	&	150	&	RTT150	&	$R_C$	&	19.131	&	0.081	\\
0.29662	&	150	&	RTT150	&	$R_C$	&	19.170	&	0.079	\\
0.29888	&	150	&	RTT150	&	$R_C$	&	19.129	&	0.077	\\
0.30113	&	150	&	RTT150	&	$R_C$	&	19.206	&	0.079	\\
0.30342	&	150	&	RTT150	&	$R_C$	&	19.183	&	0.074	\\
0.31462	&	150	&	RTT150	&	$R_C$	&	19.144	&	0.078	\\
0.31599	&	900	&	BFOSC	&	$R_C$	&	19.332	&	0.080	\\
0.31692	&	150	&	RTT150	&	$R_C$	&	19.128	&	0.075	\\
0.31917	&	150	&	RTT150	&	$R_C$	&	19.292	&	0.082	\\
0.32142	&	150	&	RTT150	&	$R_C$	&	19.349	&	0.083	\\
0.32371	&	150	&	RTT150	&	$R_C$	&	19.289	&	0.087	\\
0.32596	&	150	&	RTT150	&	$R_C$	&	19.303	&	0.088	\\
0.33725	&	150	&	RTT150	&	$R_C$	&	19.290	&	0.083	\\
0.33950	&	150	&	RTT150	&	$R_C$	&	19.275	&	0.086	\\
0.34179	&	150	&	RTT150	&	$R_C$	&	19.402	&	0.089	\\
0.34404	&	150	&	RTT150	&	$R_C$	&	19.336	&	0.089	\\
0.34629	&	150	&	RTT150	&	$R_C$	&	19.341	&	0.089	\\
0.34854	&	150	&	RTT150	&	$R_C$	&	19.454	&	0.101	\\
0.35083	&	150	&	RTT150	&	$R_C$	&	19.466	&	0.103	\\
0.35308	&	150	&	RTT150	&	$R_C$	&	19.505	&	0.104	\\
0.35533	&	150	&	RTT150	&	$R_C$	&	19.364	&	0.105	\\
0.35758	&	150	&	RTT150	&	$R_C$	&	19.383	&	0.106	\\
0.35983	&	150	&	RTT150	&	$R_C$	&	19.417	&	0.106	\\
0.36208	&	150	&	RTT150	&	$R_C$	&	19.434	&	0.107	\\
0.36433	&	150	&	RTT150	&	$R_C$	&	19.462	&	0.111	\\
0.36658	&	150	&	RTT150	&	$R_C$	&	19.270	&	0.101	\\
0.36883	&	150	&	RTT150	&	$R_C$	&	19.492	&	0.109	\\
0.37108	&	150	&	RTT150	&	$R_C$	&	19.542	&	0.122	\\
0.39230	&	600	&	DFOSC	&	$R_C$	&	19.603	&	0.051	\\
0.40064	&	600	&	DFOSC	&	$R_C$	&	19.607	&	0.039	\\
0.40800	&	600	&	DFOSC	&	$R_C$	&	19.644	&	0.037	\\
0.53319	&	600	&	DFOSC	&	$R_C$	&	19.820	&	0.044	\\
0.54030	&	600	&	DFOSC	&	$R_C$	&	19.868	&	0.041	\\
0.54841	&	600	&	DFOSC	&	$R_C$	&	19.899	&	0.042	\\
0.62960	&	600	&	DFOSC	&	$R_C$	&	20.058	&	0.090	\\
0.63685	&	300	&	DFOSC	&	$R_C$	&	20.030	&	0.142	\\
0.64096	&	300	&	DFOSC	&	$R_C$	&	20.001	&	0.145	\\
0.64468	&	300	&	DFOSC	&	$R_C$	&	19.982	&	0.158	\\
1.10432	&	540	&	Shajn 2.6m	&	$R_C$	&	20.382	&	0.120	\\
1.11983	&	900	&	RTT150	&	$R_C$	&	20.426	&	0.069	\\
1.12784	&	540	&	Shajn 2.6m	&	$R_C$	&	20.478	&	0.113	\\
1.13576	&	1080	&	Maidanak	&	$R_C$	&	20.551	&	0.041	\\
1.31530	&	900	&	BFOSC	&	$R_C$	&	20.700	&	0.133	\\
1.31725	&	900	&	RTT150	&	$R_C$	&	20.512	&	0.068	\\
1.33046	&	900	&	BFOSC	&	$R_C$	&	20.582	&	0.108	\\
1.35328	&	600	&	DFOSC	&	$R_C$	&	20.648	&	0.043	\\
1.36091	&	600	&	DFOSC	&	$R_C$	&	20.668	&	0.040	\\
1.37108	&	600	&	DFOSC	&	$R_C$	&	20.686	&	0.042	\\
1.37829	&	600	&	DFOSC	&	$R_C$	&	20.699	&	0.038	\\
1.49769	&	600	&	DFOSC	&	$R_C$	&	20.864	&	0.038	\\
1.50549	&	600	&	DFOSC	&	$R_C$	&	20.836	&	0.039	\\
1.51330	&	600	&	DFOSC	&	$R_C$	&	20.871	&	0.037	\\
1.60597	&	600	&	DFOSC	&	$R_C$	&	20.954	&	0.047	\\
1.61369	&	600	&	DFOSC	&	$R_C$	&	21.031	&	0.059	\\
1.62117	&	600	&	DFOSC	&	$R_C$	&	20.986	&	0.067	\\
2.11423	&	1260	&	Maidanak	&	$R_C$	&	21.565	&	0.100	\\
2.14083	&	900	&	RTT150	&	$R_C$	&	21.373	&	0.080	\\
2.15346	&	900	&	RTT150	&	$R_C$	&	21.330	&	0.080	\\
2.16637	&	900	&	RTT150	&	$R_C$	&	21.389	&	0.082	\\
2.29040	&	2400	&	TLS 1.34m	&	$R_C$	&	21.700	&	0.130	\\
2.32674	&	900	&	MOSCA	&	$R_C$	&	21.671	&	0.188	\\
2.33858	&	900	&	RTT150	&	$R_C$	&	21.619	&	0.091	\\
2.35204	&	900	&	RTT150	&	$R_C$	&	21.587	&	0.083	\\
2.36504	&	900	&	RTT150	&	$R_C$	&	21.875	&	0.115	\\
3.10104	&	1800	&	Maidanak	&	$R_C$	&	22.286	&	0.060	\\
3.16792	&	3000	&	RTT150	&	$R_C$	&	22.062	&	0.072	\\
3.25538	&	3000	&	RTT150	&	$R_C$	&	22.239	&	0.077	\\
3.28438	&	1800	&	MOSCA	&	$R_C$	&	22.213	&	0.052	\\
3.34992	&	3000	&	RTT150	&	$R_C$	&	22.338	&	0.095	\\
3.73440	&	660	&	Keck/LRIS	&	$R_C$	&	22.758	&	0.042	\\
4.11354	&	2700	&	Maidanak	&	$R_C$	&	23.181	&	0.100	\\
4.23200	&	7200	&	RTT150	&	$R_C$	&	23.072	&	0.108	\\
4.37326	&	1800	&	MOSCA	&	$R_C$	&	23.244	&	0.099	\\
4.87336	&	900	&	Keck/LRIS	&	$R_C$	&	23.587	&	0.123	\\
5.10382	&	3420	&	Maidanak	&	$R_C$	&	23.733	&	0.096	\\
5.34483	&	9000	&	RTT150	&	$R_C$	&	23.680	&	0.201	\\
5.35173	&	3600	&	ALFOSC	&	$R_C$	&	23.571	&	0.088	\\
6.14534	&	3420	&	Maidanak	&	$R_C$	& $>	23.6	$ &	$\cdots$	\\
7.17834	&	3600	&	Maidanak	&	$R_C$	& $>	23.8	$ &	$\cdots$	\\
7.26820	&	12000	&	TLS 1.34m	&	$R_C$	& $>	23.7	$ &	$\cdots$	\\
7.37882	&	3600	&	ALFOSC	&	$R_C$	&	24.602	&	0.039	\\
10.3198	&	3600	&	ALFOSC	&	$R_C$	& $>	24.6	$ &	$\cdots$	\\
272.654	&	2500	&	FORS2	&	$R_C$	& $>	27.1	$ &	$\cdots$	\\
672.332	&	7500	&	FORS2	&	$R_C$	&	{\it (combined)}	&	$\cdots$	\\
\hline									
0.0010848	&	10	&	Watcher	&	$CR$	&	16.500	&	0.400	\\
0.0022314	&	10	&	Watcher	&	$CR$	&	16.160	&	0.220	\\
0.0033890	&	10	&	Watcher	&	$CR$	&	16.460	&	0.220	\\
0.0046149	&	60	&	Watcher	&	$CR$	&	16.830	&	0.380	\\
0.0088341	&	60	&	Watcher	&	$CR$	&	17.090	&	0.450	\\
0.0130493	&	60	&	Watcher	&	$CR$	&	17.230	&	0.440	\\
0.0172518	&	60	&	Watcher	&	$CR$	&	17.110	&	0.360	\\
0.0214654	&	60	&	Watcher	&	$CR$	&	17.430	&	0.440	\\
0.0263665	&	120	&	Watcher	&	$CR$	&	17.420	&	0.320	\\
0.0328664	&	38.6	&	ROTSE	&	$CR$	& $>	17.412	$ &	$\cdots$	\\
0.0340538	&	120	&	Watcher	&	$CR$	&	17.620	&	0.370	\\
0.0351278	&	143.89	&	ROTSE	&	$CR$	&	17.625	&	0.192	\\
0.0385665	&	143.28	&	ROTSE	&	$CR$	&	17.570	&	0.170	\\
0.0417402	&	120	&	Watcher	&	$CR$	&	17.590	&	0.330	\\
0.0419979	&	143.28	&	ROTSE	&	$CR$	&	17.558	&	0.147	\\
0.0454286	&	143.23	&	ROTSE	&	$CR$	&	18.054	&	0.206	\\
0.0488545	&	143.18	&	ROTSE	&	$CR$	&	17.792	&	0.164	\\
0.0494170	&	120	&	Watcher	&	$CR$	&	17.783	&	0.090	\\
0.0522827	&	143.23	&	ROTSE	&	$CR$	&	17.785	&	0.153	\\
0.0548713	&	67.49	&	ROTSE	&	$CR$	&	17.927	&	0.321	\\
0.0571028	&	120	&	Watcher	&	$CR$	&	17.810	&	0.090	\\
0.0574228	&	143.84	&	ROTSE	&	$CR$	&	17.723	&	0.155	\\
0.0608539	&	143.43	&	ROTSE	&	$CR$	&	17.624	&	0.139	\\
0.0639448	&	113.58	&	ROTSE	&	$CR$	&	17.841	&	0.188	\\
0.0647885	&	120	&	Watcher	&	$CR$	&	17.883	&	0.100	\\
0.0662157	&	38.6	&	ROTSE	&	$CR$	&	17.659	&	0.282	\\
0.0684630	&	148	&	ROTSE	&	$CR$	&	17.876	&	0.200	\\
0.0724741	&	120	&	Watcher	&	$CR$	&	17.850	&	0.090	\\
0.0742371	&	343.17	&	ROTSE	&	$CR$	&	17.780	&	0.088	\\
0.0801480	&	120	&	Watcher	&	$CR$	&	17.700	&	0.080	\\
0.0822853	&	342.81	&	ROTSE	&	$CR$	&	17.905	&	0.074	\\
0.0878566	&	120	&	Watcher	&	$CR$	&	17.790	&	0.090	\\
0.0903340	&	342.96	&	ROTSE	&	$CR$	&	17.959	&	0.064	\\
0.0955536	&	120	&	Watcher	&	$CR$	&	17.887	&	0.090	\\
0.0983836	&	342.71	&	ROTSE	&	$CR$	&	18.049	&	0.091	\\
0.1032274	&	120	&	Watcher	&	$CR$	&	18.070	&	0.100	\\
0.1064264	&	342.4	&	ROTSE	&	$CR$	&	18.114	&	0.081	\\
0.1109128	&	120	&	Watcher	&	$CR$	&	17.890	&	0.090	\\
0.1179262	&	335.45	&	ROTSE	&	$CR$	&	18.141	&	0.079	\\
0.1185981	&	120	&	Watcher	&	$CR$	&	18.210	&	0.120	\\
0.1262346	&	342.96	&	ROTSE	&	$CR$	&	18.232	&	0.078	\\
0.1262834	&	120	&	Watcher	&	$CR$	&	18.073	&	0.100	\\
0.1339918	&	120	&	Watcher	&	$CR$	&	18.113	&	0.110	\\
0.1342816	&	342.96	&	ROTSE	&	$CR$	&	18.255	&	0.069	\\
0.1416771	&	120	&	Watcher	&	$CR$	&	18.323	&	0.140	\\
0.1423343	&	343.47	&	ROTSE	&	$CR$	&	18.354	&	0.066	\\
0.1493624	&	120	&	Watcher	&	$CR$	&	18.170	&	0.120	\\
0.1503864	&	343.01	&	ROTSE	&	$CR$	&	18.407	&	0.070	\\
0.1584325	&	342.25	&	ROTSE	&	$CR$	&	18.440	&	0.070	\\
0.1611113	&	975	&	Watcher	&	$CR$	&	18.423	&	0.040	\\
0.1682310	&	342.61	&	ROTSE	&	$CR$	&	18.484	&	0.180	\\
0.1724372	&	975	&	Watcher	&	$CR$	&	18.560	&	0.050	\\
0.1762647	&	342.3	&	ROTSE	&	$CR$	&	18.666	&	0.093	\\
0.1843071	&	343.27	&	ROTSE	&	$CR$	&	18.578	&	0.082	\\
0.1844514	&	1098	&	Watcher	&	$CR$	&	18.540	&	0.050	\\
0.1923501	&	342.46	&	ROTSE	&	$CR$	&	18.541	&	0.070	\\
0.1971843	&	1097	&	Watcher	&	$CR$	&	18.517	&	0.050	\\
0.2003860	&	342.56	&	ROTSE	&	$CR$	&	18.637	&	0.075	\\
0.2084307	&	342.61	&	ROTSE	&	$CR$	&	18.698	&	0.096	\\
0.2106114	&	1220	&	Watcher	&	$CR$	&	18.707	&	0.060	\\
0.2164729	&	342.96	&	ROTSE	&	$CR$	&	18.782	&	0.101	\\
0.2245117	&	342.41	&	ROTSE	&	$CR$	&	18.789	&	0.102	\\
0.2254456	&	1342	&	Watcher	&	$CR$	&	18.847	&	0.060	\\
0.2325434	&	342.35	&	ROTSE	&	$CR$	&	18.881	&	0.084	\\
0.2405816	&	342.46	&	ROTSE	&	$CR$	&	18.941	&	0.123	\\
0.2410154	&	1343	&	Watcher	&	$CR$	&	18.753	&	0.040	\\
0.2486180	&	342.51	&	ROTSE	&	$CR$	&	18.877	&	0.105	\\
0.2566562	&	342.82	&	ROTSE	&	$CR$	&	18.958	&	0.095	\\
0.2572685	&	1464	&	Watcher	&	$CR$	&	19.033	&	0.050	\\
0.2647067	&	343.17	&	ROTSE	&	$CR$	&	18.888	&	0.170	\\
0.2727542	&	342.96	&	ROTSE	&	$CR$	&	18.954	&	0.109	\\
0.2749340	&	1587	&	Watcher	&	$CR$	&	18.987	&	0.050	\\
0.2807917	&	342.3	&	ROTSE	&	$CR$	&	18.995	&	0.118	\\
0.2888233	&	342.46	&	ROTSE	&	$CR$	&	19.122	&	0.132	\\
0.2940183	&	1709	&	Watcher	&	$CR$	&	19.053	&	0.060	\\
0.2968695	&	342.96	&	ROTSE	&	$CR$	&	19.125	&	0.210	\\
0.3049105	&	342.61	&	ROTSE	&	$CR$	&	19.108	&	0.169	\\
0.3129479	&	342.56	&	ROTSE	&	$CR$	&	19.318	&	0.159	\\
0.3145154	&	1831	&	Watcher	&	$CR$	&	19.397	&	0.110	\\
0.3209870	&	342.45	&	ROTSE	&	$CR$	&	19.300	&	0.132	\\
0.3290197	&	342.4	&	ROTSE	&	$CR$	&	19.327	&	0.173	\\
0.3330194	&	689.53	&	ROTSE	&	$CR$	&	19.360	&	0.174	\\
0.3364198	&	1952	&	Watcher	&	$CR$	&	19.400	&	0.100	\\
0.3370540	&	342.5	&	ROTSE	&	$CR$	&	19.305	&	0.163	\\
0.3450996	&	342.86	&	ROTSE	&	$CR$	&	19.551	&	0.213	\\
0.3531434	&	342.45	&	ROTSE	&	$CR$	&	19.353	&	0.196	\\
0.3597482	&	2075	&	Watcher	&	$CR$	&	19.683	&	0.100	\\
0.3651768	&	689.38	&	ROTSE	&	$CR$	&	19.464	&	0.201	\\
0.3812533	&	690.3	&	ROTSE	&	$CR$	&	19.480	&	0.220	\\
0.3844837	&	2197	&	Watcher	&	$CR$	&	19.777	&	0.100	\\
0.3982605	&	770.22	&	ROTSE	&	$CR$	& $>	19.741	$ &	$\cdots$	\\
0.4241566	&	1462.69	&	ROTSE	&	$CR$	&	19.687	&	0.252	\\
1.0566720	&	1037.58	&	ROTSE	&	$CR$	& $>	19.602	$ &	$\cdots$	\\
1.0986892	&	8554	&	Watcher	&	$CR$	&	20.397	&	0.070	\\
1.1985530	&	8688	&	Watcher	&	$CR$	&	20.343	&	0.080	\\
1.3005326	&	8923	&	Watcher	&	$CR$	&	20.550	&	0.140	\\
2.1764170	&	22125	&	Watcher	&	$CR$	&	21.507	&	0.170	\\
2.4570745	&	1040.76	&	ROTSE	&	$CR$	& $>	19.937	$ &	$\cdots$	\\
\hline
0.20221	&	180	&	MIRO	&	$I_C$	&	18.311	&	0.149	\\
0.26296	&	900	&	BFOSC	&	$I_C$	&	18.549	&	0.070	\\
0.30365	&	900	&	BFOSC	&	$I_C$	&	18.715	&	0.055	\\
0.34481	&	900	&	BFOSC	&	$I_C$	&	18.901	&	0.070	\\
0.35175	&	360	&	ANDICAM	&	$I_C$	&	18.940	&	0.090	\\
0.35683	&	360	&	ANDICAM	&	$I_C$	&	18.937	&	0.062	\\
0.36191	&	360	&	ANDICAM	&	$I_C$	&	19.052	&	0.056	\\
0.36692	&	360	&	ANDICAM	&	$I_C$	&	18.967	&	0.043	\\
0.37107	&	600	&	DFOSC	&	$I_C$	&	19.009	&	0.048	\\
0.37200	&	360	&	ANDICAM	&	$I_C$	&	18.987	&	0.036	\\
0.37888	&	600	&	DFOSC	&	$I_C$	&	19.050	&	0.045	\\
0.38632	&	600	&	DFOSC	&	$I_C$	&	19.118	&	0.042	\\
0.47240	&	360	&	ANDICAM	&	$I_C$	&	19.266	&	0.059	\\
0.47737	&	360	&	ANDICAM	&	$I_C$	&	19.310	&	0.050	\\
0.48238	&	360	&	ANDICAM	&	$I_C$	&	19.338	&	0.052	\\
0.48747	&	360	&	ANDICAM	&	$I_C$	&	19.361	&	0.060	\\
0.49254	&	360	&	ANDICAM	&	$I_C$	&	19.359	&	0.048	\\
1.38701	&	600	&	DFOSC	&	$I_C$	&	20.152	&	0.039	\\
1.39420	&	600	&	DFOSC	&	$I_C$	&	20.294	&	0.042	\\
1.40178	&	600	&	DFOSC	&	$I_C$	&	20.223	&	0.049	\\
2.37793	&	900	&	MOSCA	&	$I_C$	&	21.156	&	0.041	\\
\hline
0.36436	&	360	&	ANDICAM	&	$J$	&	18.090	&	0.080	\\
0.48350	&	1177	&	PAIRITEL	&	$J$	&	18.638	&	0.054	\\
0.48494	&	360	&	ANDICAM	&	$J$	&	18.580	&	0.090	\\
0.51112	&	2260	&	PAIRITEL	&	$J$	&	19.196	&	0.114	\\
0.56496	&	683	&	PAIRITEL	&	$J$	&	19.472	&	0.141	\\
0.62604	&	2190	&	PAIRITEL	&	$J$	&	18.872	&	0.054	\\
1.53323	&	706	&	PAIRITEL	&	$J$	&	19.575	&	0.138	\\
1.61412	&	2237	&	PAIRITEL	&	$J$	&	19.725	&	0.157	\\
2.36277	&	360	&	ANDICAM	&	$J$	& $>	19.1	$ &	$\cdots$	\\
2.52278	&	2237	&	PAIRITEL	&	$J$	& $>	20.1	$ &	$\cdots$	\\
3.46425	&	2213	&	PAIRITEL	&	$J$	& $>	20.2	$ &	$\cdots$	\\
4.50410	&	2266	&	PAIRITEL	&	$J$	& $>	20.3	$ &	$\cdots$	\\
5.52048	&	2260	&	PAIRITEL	&	$J$	& $>	20.3	$ &	$\cdots$	\\
\hline									
0.48350	&	1177	&	PAIRITEL	&	$H$	&	17.693	&	0.141	\\
0.51112	&	2260	&	PAIRITEL	&	$H$	&	17.784	&	0.127	\\
0.56496	&	683	&	PAIRITEL	&	$H$	&	17.890	&	0.136	\\
5.52048	&	2260	&	PAIRITEL	&	$H$	& $>	19.7	$ &	$\cdots$	\\
0.62604	&	2190	&	PAIRITEL	&	$H$	&	17.978	&	0.136	\\
1.53323	&	706	&	PAIRITEL	&	$H$	& $>	19.2	$ &	$\cdots$	\\
1.61412	&	2237	&	PAIRITEL	&	$H$	&	19.090	&	0.278	\\
2.52278	&	2237	&	PAIRITEL	&	$H$	& $>	19.6	$ &	$\cdots$	\\
3.50562	&	2237	&	PAIRITEL	&	$H$	& $>	19.8	$ &	$\cdots$	\\
4.54525	&	2260	&	PAIRITEL	&	$H$	& $>	19.8	$ &	$\cdots$	\\
\hline
0.48350	&	1177	&	PAIRITEL	&	$K_S$	&	16.913	&	0.174	\\
0.51112	&	2260	&	PAIRITEL	&	$K_S$	&	17.162	&	0.217	\\
0.56496	&	683	&	PAIRITEL	&	$K_S$	&	17.197	&	0.185	\\
0.62604	&	2190	&	PAIRITEL	&	$K_S$	&	17.326	&	0.191	\\
1.53323	&	706	&	PAIRITEL	&	$K_S$	& $>	18	$ &	$\cdots$	\\
1.61412	&	2237	&	PAIRITEL	&	$K_S$	&	18.148	&	0.265	\\
2.52278	&	2237	&	PAIRITEL	&	$K_S$	& $>	18.8	$ &	$\cdots$	\\
3.46425	&	2213	&	PAIRITEL	&	$K_S$	& $>	18.8	$ &	$\cdots$	\\
4.54525	&	2260	&	PAIRITEL	&	$K_S$	& $>	18.9	$ &	$\cdots$	\\
5.52048	&	2260	&	PAIRITEL	&	$K_S$	& $>	19.0	$ &	$\cdots$	\\
\hline									
1169.785	&	7864	&	HST ACS	&	$F775W_{AB}$	& $>	28.5$ &	$\cdots$	\\
\hline									
\hline									
\end{longtable}
\tablefoot{UVOT is the 30 cm UltraViolet and Optical Telescope onboard the \emph{Swift} satellite.\\
Maidanak is the 1.5m telescope of the Maidanak observatory in Uzbekistan.\\
RTT150 is the 1.5m Russian-Turkish Telescope at T\"UB\.ITAK National Observatory on Mount Bakyrlytepe, Antalya, Turkey.\\
MOSCA is the MOSaic CAmera on the 2.5m Nordic Optical Telescope, La Palma, Canary Islands, Spain.\\
Keck/LRIS is the Low Resolution Imaging Spectrograph on the 10m Keck I telescope, Mauna Kea, Hawaii, United States of America.\\
DOLORES is the Device Optimized for the LOw RESolution detector on the 3.6m TNG (Telescopio Nazionale Galileo) telescope on La Palma, Canary Islands, Spain.\\
MIRO is the 1.2m telescope of the Mt. Abu Infrared Observatory, India.\\
BFOSC is the Bologna Faint Object Spectrograph \& Camera at the G. D. Cassini 152 cm  telescope of the Bologna University, Loiano, Italy.\\
DFOSC is the Danish Faint Object Spectrograph and Camera on the Danish 1.54m telescope on La Silla, Chile.\\
Watcher is the 0.3m Watcher robotic telescope at Boyden Observatory, South Africa.\\
Shajn is the 2.6m Shajn telescope of the Crimean Astrophysical Observatory (CrAO), Ukraine.\\
TLS 1.34m is the 1.34m Schmidt telescope of the Th\"uringer Landessternwarte Tautenburg, Germany.\\
ALFOSC is the Andalucia Faint Object Spectrograph and Camera on the 2.5m Nordic Optical Telescope, La Palma, Canary Islands, Spain.\\
FORS2 is the FOcal Reducer and low dispersion Spectrograph 2 on the 8.2m Very Large Telescope, Paranal Observatory, Chile.\\
ROTSE is the 0.3m Robotic Optical Transient Search Experiment III-B telescope at the H.E.S.S. site, Mt. Gamsberg, Namibia.\\
ANDICAM is the A Novel Double-Imaging CAMera detector on the 1.3m Small and Moderate Aperture Research Telescope System (SMARTS) telescope at the Cerro Tololo Interamerican Observatory in Chile.\\
PAIRITEL is the 1.3m Peters Automatic InfraRed Imaging TELescope on Mt. Hopkins, Arizona, United States of America.\\
HST ACS is the Hubble Space Telescope equipped with the Advanced Camera for Surveys.}\\
}
}

In order to get a good coverage of the light curve, we obtained data using several different telescopes around the world. Our complete data set comprises 
a total of 412 points from the UV to K-band, one of the largest photometric samples of an optical/NIR afterglow in the \emph{Swift} era.

The earliest dataset was obtained by the Watcher telescope, located at Boyden Observatory, South Africa, starting 36.2 s after the burst, followed by \emph{Swift} UVOT starting 86 s after the trigger. Early ground based optical data were obtained with the ROTSE-IIIc 0.3m telescope at the H.E.S.S. site at Mt. Gamsberg, Namibia, the 1.5m telescope on Mt. Maidanak/Uzbekistan, the 2.6m Shajn telescope at CrAO (Crimean Astrophysical Observatory/Ukraine), the TNG (Telescopio Nazionale Galileo) on La Palma equipped with DOLoRes, the 1.2m MIRO telescope on Mt. Abu/India, with BFOSC (Bologna Faint Object Spectrograph \& Camera) at the G. D. Cassini 152 cm  telescope of the Bologna University under poor conditions and with the RTT150 (1.5m Russian-Turkish telescope, Bakirlitepe, Turkey), the RTT150 data are also presented in \cite{Khamitov08}. 

\begin{figure*}
  \centering
 \includegraphics[width=\textwidth]{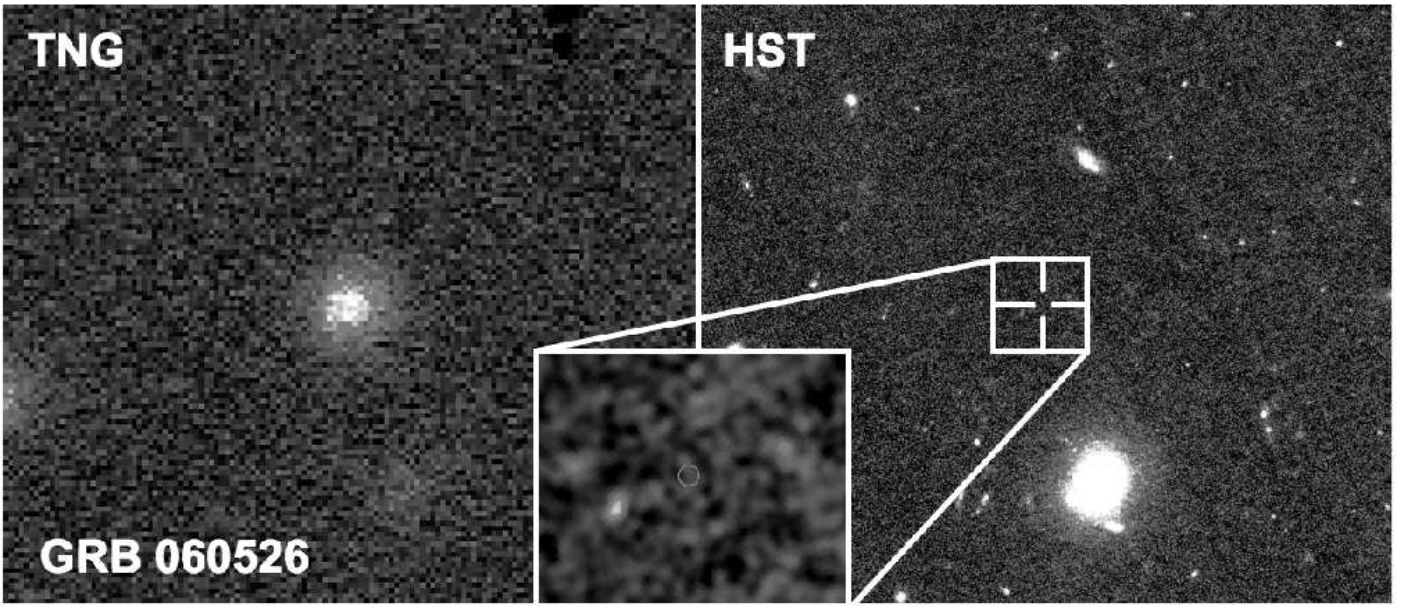}
\caption{Left panel: An image from the TNG with the afterglow present. Right panel: The same region observed with \emph{HST}/ACS and a zoom-in around the region of the afterglow. There is no source detected at the afterglow position itself down to a limit of F775W(AB)$>28.5$. The brighter spot to the South-East in the zoomed {\it HST} image corresponds to the host candidate from the VLT mentioned in the text.}
              \label{hostfig}
    \end{figure*}

The light curve was followed up sparsely every night over nearly a week with DFOSC (Danish Faint Object Spectrograph and Camera) on the Danish 1.54m telescope on La Silla/Chile under partially photometric conditions and with MOSCA (MOSaic CAmera) and ALFOSC (Andalucia Faint Object Spectrograph and Camera) at the Nordic Optical Telescope on La Palma. Furthermore, two epochs were obtained with the Tautenburg 1.34m Schmidt telescope and two sets of images were taken several days after the GRB with Keck/LRIS simultaneously in the Kron-Cousins $R$ and the Sloan $g^\prime$ bands, the second observation was performed at high airmass under bad seeing conditions. Late images were obtained with FORS2 at the VLT on Paranal/Chile on Feb. 23, 2007 and Mar. 30, 2008
 in the $R_C$ band with exposure times of 2500 and 7500 s, respectively, to look for the host galaxy. We also took observations with the {\em Hubble Space Telescope} on 9 August 2009, utilizing the Advanced Camera for Surveys with the F775W (roughly SDSS $i^\prime$) filter (see Fig. \ref{hostfig}). A total of 7844 s of observations were obtained in six dithered exposures. These were reduced via {\tt multidrizzle} in the standard fashion.

Near infrared data were collected with ANDICAM and the 1.3 m SMARTS telescope (Small and Moderate Aperture Research Telescope System) at CTIO under non-photometric conditions as well as with the robotic 1.3m PAIRITEL telescope on Mt. Hopkins. 

The UVOT data were reduced and analysed using the standard UVOT tasks within the {\tt heasoft} package. For the photometric calibration of the ground-based data, we determined the calibrated magnitude of six comparison stars in the field using photometric zero points from DFOSC in the $V$, $R_C$ and $I_C$ bands (see Table \ref{compstars}). These stars were then used to perform relative Point-Spread Function (PSF) photometry to get the calibrated magnitude of the afterglow. For some of the late NOT images as well as the faint MIRO detections, though, we applied relative aperture photometry using a circle of 20 pixels diameter (and an annulus of 10 pixels for the sky). For the $B$-band, where no DFOSC data were available, we took zero points for only three comparison stars from the SDSS (see Table \ref{compstars}), converting them with the equations of \cite{JesterSDSS}. $g^\prime$ zero points were taken from the SDSS, and they are given as AB magnitudes. For the RTT 150 data, we used one USNO-B1 star as reference that was calibrated using Landolt standard stars. The results are in full agreement with the rest of the data set. The Watcher data were analysed using a dedicated photometry pipeline \citep{Ferrero2010}. The $J$, $H$ and $K_S$ band data from SMARTS/ANDICAM and PAIRITEL were calibrated using three and ten nearby stars, respectively, from the 2MASS catalogue. For the $H$ and $K_S$ band data, we used smaller apertures and applied aperture corrections to reduce the influence of the highly variable background.

\begin{figure*}
  \centering
  \includegraphics[width=18cm]{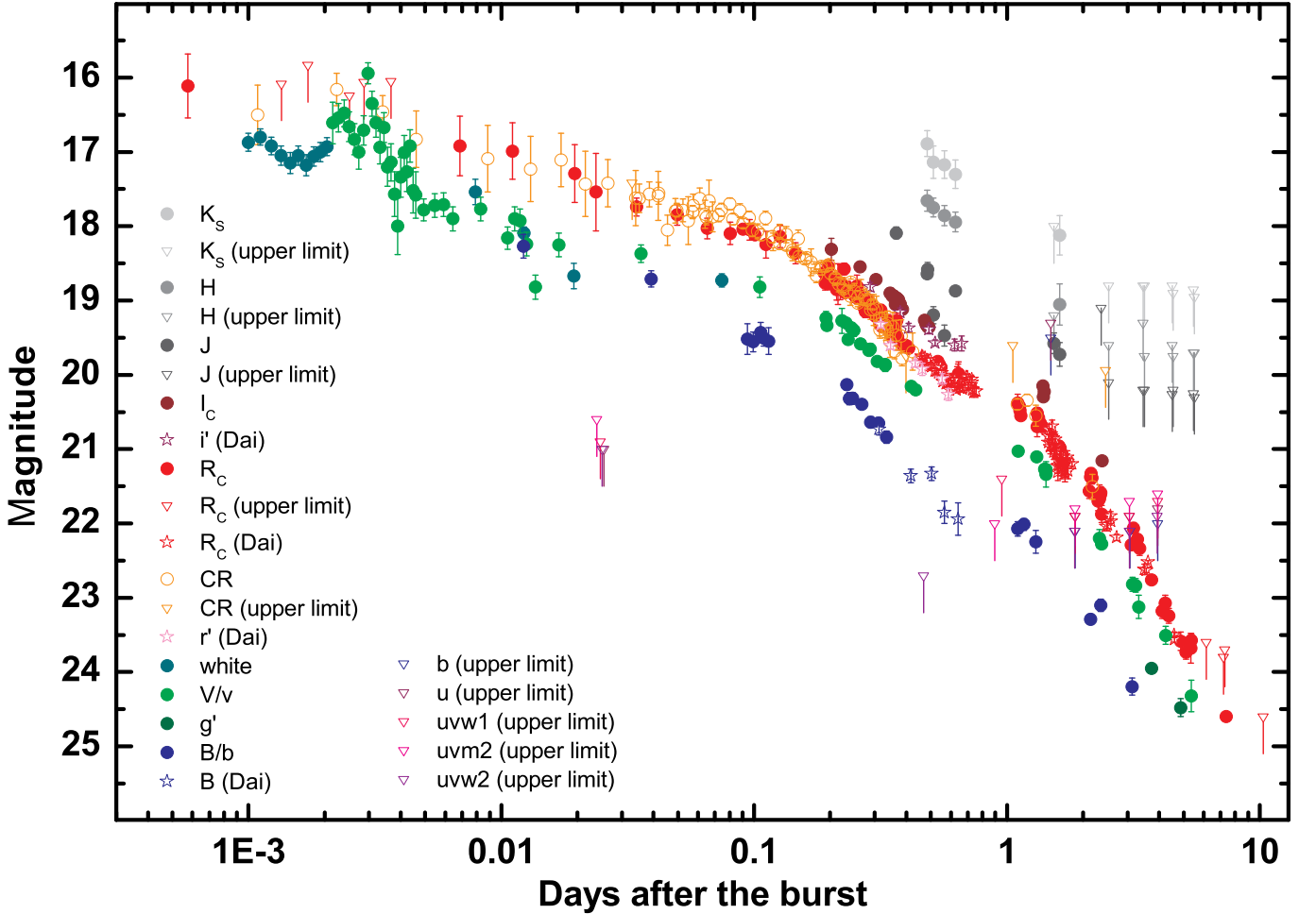}
   \caption{Light curves of the optical afterglow of GRB 060526. Filled and open circles (detections, $Bg^\prime VR_C\,CR\,I_CJHK_S$) and open triangles (upper limits, $uvw2\,uvm2\,uvw1\,ubR_C\,CR\,JHK_S$) are presented in this work, and open stars ($Br^\prime Ri^\prime$) are taken from \cite{Dai2007}. The early light curve especially shows strong variability, with two flares seen in the UVOT $v$ band data. The later light curve also shows evident variability, and the large $V-R$, $B-R$ colours due to the high redshift are evident. Data in this plot are not corrected for foreground extinction.}
              \label{lightcurve}
    \end{figure*}

All data and upper limits are given in Table \ref{obslog}, the data are not corrected for Galactic extinction. Note we give the $g^\prime$ and HST $F775W$ magnitudes in AB magnitudes.
For the final light curve fitting, we add $Br^\prime R_Ci^\prime$ band data from \cite{Dai2007}. We shift the $B$ and $R_C$ band data of \cite{Dai2007} by 0.1 magnitudes to bring it to our zero point. The multi-color light curves are shown in Fig. \ref{lightcurve}.

\subsection{Spectroscopy}
Spectra were obtained with FORS1/VLT on May 27 from 9 to 12 hours after the burst. Four different grisms cover the wavelength range from 3650 to 9200 {\AA}. For all four grisms a 1$\farcs$0 slit was used which provides resolutions between 2.4 and 11.1 {\AA}. Reduction, cosmic ray removal, extraction and wavelength calibration were performed using standard tasks in IRAF\footnote{http://iraf.noao.edu}. The final spectra were then normalised as no absolute flux calibration was needed. In order to improve the S/N, we combined the datasets taken with the same grism weighted with their variance. A summary of the spectroscopic observations is given in Table~\ref{speclog}.

\begin{table*}
\caption{Spectra of GRB 060526 from FORS1/VLT. Times given are the start time of the observations, midexposure times are in days after the GRB trigger.}
\label{speclog}    
\centering                        
\begin{tabular}{l c c c c c c}       
\hline\hline   
Time (May & Midtime & Exptime & Grism & Spectral range& Resolution\\
27 [UT]) & [d] & [s] & &  [\AA] &[\AA] \\
\hline
01:16:56 & 0.37042 & 600& 300V & 3650 -- 8900 & 11.1\\
01:29:46 & 0.38105 & 900 & 600I & 7000 -- 9200 & 4.4\\
01:47:28 & 0.39335 & 900 & 1200B & 3860 -- 4400 & 2.4\\
02:05:27 & 0.40583 & 900 & 600V & 4080 -- 7200 & 4.5\\
02:24:00 & 0.42043 & 1200 & 300V & 3650 -- 8900 & 11.1\\
02:57:05 & 0.44681 & 1800 & 1200B & 3860 -- 4400 & 2.4\\
03:30:34 & 0.47007 & 1800 & 600V & 4080 -- 7200 & 4.5\\
04:04:08 & 0.49339 & 1800 & 600I & 7000 -- 9200 & 4.4\\
\hline \hline
\end{tabular}
\end{table*}

\begin{table*}
\caption{Magnitudes of the comparison stars used for photometry.}
\label{compstars}    
\centering                        
\begin{tabular}{l l l c c c c}       
\hline\hline   
\# &Coordinates && $B$ & $V$ & $R_C$ & $I_C$ \\
&RA &Dec.&[mag]&[mag]&[mag]&[mag]\\ \hline
1     &15:31:19.6  &+00:16:59.5&$19.90\pm0.03$&$19.36\pm0.02$&$19.02\pm0.02$&$18.64\pm0.02$\\
2     &15:31:22.7  &+00:17:21.2&$20.30\pm0.03$&$19.51\pm0.02$&$19.09\pm0.02$&$18.71\pm0.02$\\
3     &15:31:20.9  &+00:16:39.1&$19.97\pm0.03$&$19.11\pm0.02$&$18.55\pm0.02$&$18.12\pm0.02$\\
4     &15:31:14.4  &+00:17:49.9&---&           $19.78\pm0.02$&$19.43\pm0.02$&$18.81\pm0.03$\\
5     &15:31:16.5  &+00:17:55.5&---&           $20.34\pm0.03$&$19.96\pm0.03$&$20.59\pm0.03$\\
6     &15:31:18.3  &+00:18:17.7&---&           $20.49\pm0.03$&$19.45\pm0.02$&$18.23\pm0.02$\\ \hline \hline
\end{tabular}
\end{table*}


\section{Prompt emission}
Up to two hours after the burst, the light curve features an optical flare contemporaneous to the XRT/BAT flare at $\approx250$ s and a following plateau phase \citep{BrownGCN, FrenchGCN, RykoffGCN}.  We reduced the BAT event data using standard procedures within the software provided by {\tt HEASOFT} (version 6.1). The XRT observations were reduced using the standard {\tt xrtpipeline} (version 0.10.4) for XRT data analysis software using the most recent calibration files. The spectral data of BAT and XRT were analysed with {\tt XSPEC} version 11.3 \citep{Arnaud96}. The X-ray Galactic column density was fixed to $5.02\times10^{20}$ cm$^{-2}$ \citep{Kalberla05}. We estimated the late-time extragalactic column density by fitting the XRT PC data from 517 s to $1.34\times10^{5}$ s post trigger, where spectral evolution is negligible, using an absorbed power law model. We only find an upper limit of N$_\mathrm{HX} < 9.8\times10^{21}$ cm$^{-2}$ \cite[see also][]{Campana10}.

\begin{figure}
  \centering
 \includegraphics[width=\columnwidth]{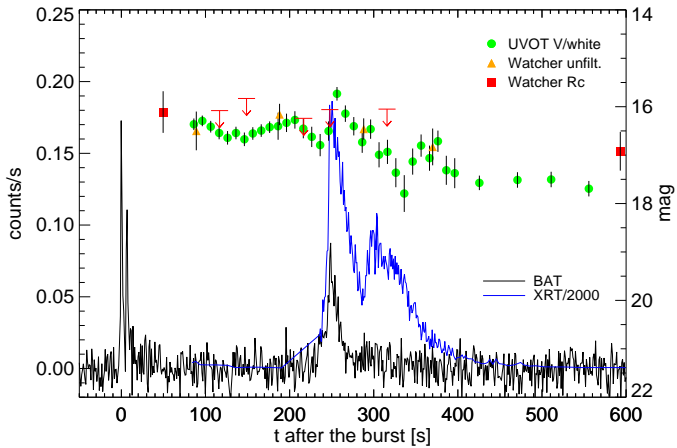}
   \caption{The two early flares seen in the BAT and XRT data and temporally coincident optical observations by UVOT and Watcher. The second flare is much softer at high energies and hence it is only seen in the XRT data. Both flares are reflected in the optical data with the first optical flare being coincident with the first BAT/XRT flare while the second flare is delayed compared to the second X-ray flare. The \emph{white} data have been shifted upward by 0.3 mag to the $v$ zero point.}
              \label{060526:flare}
    \end{figure}

In Fig. \ref{060526:flare} we compare the timescale of the flares at high energies (BAT and XRT) with contemporaneous optical data. The first, stronger flare is seen both by the XRT and BAT whereas the second, softer flare is only visible in the XRT data. Our earliest optical data are also coincident with the two flares. A broad early bump is observed in the optical light curve which precedes the second episode of BAT emission and the major X-ray pulse. The Watcher data only show a plateau during the high-energy flares due to a relatively low time resolution. The higher time-resolution of the UVOT data shows that there are also two significant flares in the optical. The first optical flare is contemporaneous with the peak of the BAT/XRT flare within $2\sigma$ of the temporal error. Then, there may also be a small optical bump (significance only $\approx1.5\sigma$) at the time of the second (XRT-only) flare, whereas the second significant optical flare occurs $\approx90$ s after the XRT-only flare implying that these two events are probably not connected. Power-law fitting of the optical light curve (with respect to the BAT trigger time T$_0$) shows the slopes are very steep. The rising slope of the first flare is $\alpha_{r1}=-12.8\pm3.0$, and the decay slope is $\alpha_{d1}=5.8\pm0.8$, or, if one takes the bump as a second flare, the two decay slopes are $\alpha_{d11}=8.2\pm2.0$, $\alpha_{d12}=8.6\pm2.5$. The rise and decay slopes for the second peak are similarly steep, but with larger uncertainties.
 
Prompt optical flashes attributed to reverse shocks, as seen in the landmark burst GRB 990123 \citep{Akerlof99}, have been observed in only a few cases since the launch of Swift, e.g., GRB 060111B \citep{Klotz06} and GRB 060117 (\citealt{Jelinek06}, see \citealt{Kann07} for a recent overview of light curves with probable reverse shock flashes/steep decays in the \emph{Swift} era). Most bursts for which early optical data are available display no evidence of reverse shock emission. In a number of cases, the optical light curve is dominated by forward shock emission from very early times, e.g. GRB 050401 \citep{Rykoff05}, GRB 060418 and GRB 060607A \citep{Molinari07, Nysewander07}, GRB 060605 \citep{Ferrero08}, GRB 061007 \citep{Mundell07} and GRB 081008 \citep{Yuan10}. An optical component of the emission from internal shocks has been invoked to explain the correlation between the optical and high-energy light curves observed in several bursts, e.g. GRB 041219A \citep{Vestrand041219A, Blake041219A, Fan05}, GRB 050820A \citep{Vestrand050820A}, and GRB 050904 \citep{Wei06}. The very rapid variability as measured by the steep slopes and the contemporaneous first flare indicates that GRB 060526 is another case where the early optical light curve is dominated by central engine activity. Such time resolution and coincident optical and high-energy observations are still rare, however, a thorough analysis of the flares and their spectral properties goes beyond the scope of this paper.

 \section{Afterglow photometry and light curve modelling}    
    
\subsection{The multi-color light curve}
\label{MCLC}

\begin{figure}
  \centering
  \includegraphics[width=\columnwidth]{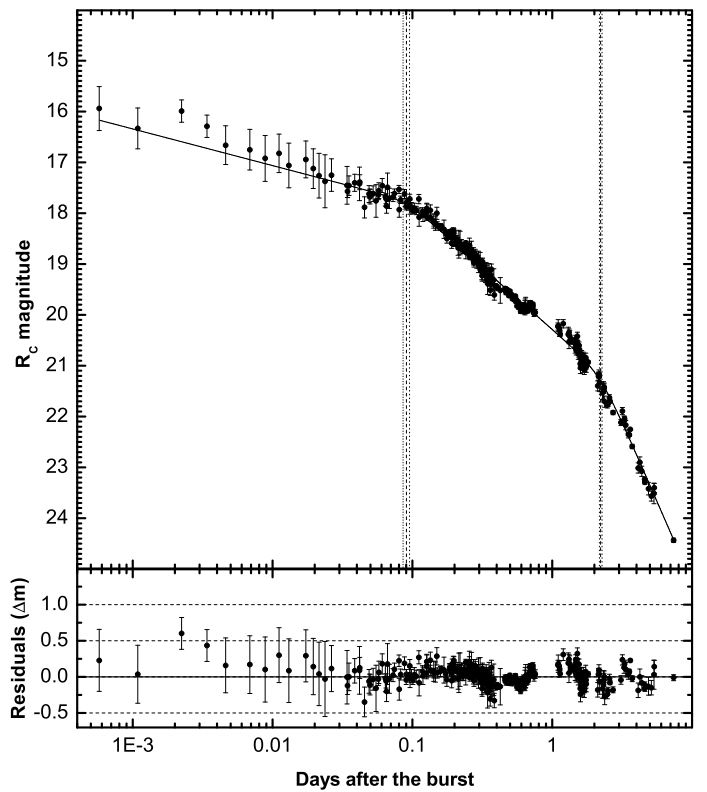}
   \caption{The $R_C$-band light curve of GRB 060526 fitted with a double smoothly-broken power law, using data starting at 0.046 days. The residuals of the fit show an early optical flare and later strong variations of up to 0.3 magnitudes. The dashed vertical lines mark the break times, the dotted lines the $1\sigma$ region of uncertainty. The second break is soft ($n\approx4$) and is thus seen as a smooth rollover.}
              \label{power-law}
    \end{figure}

To analyse the evolution of the light curve, we use the $R_C/CR$ band which has the densest sampling, a total of 319 data points in all, from 50 s to more than 7 days after the GRB. For all fits, we fix the host magnitude to $m_h=29$ (see \kref{Host}). A fit with a single power law to these data is very strongly rejected, with $\csr31.89$ (with 319 d.o.f.). Even if we remove the earliest data which are affected by optical flares, starting only at 400 s after the GRB, the fit is still rejected ($\csr29.46$ with 308 d.o.f.). A double smoothly-broken power law gives a much better fit, but even so, the fit is formally rejected, with $\csr3.12$ with 302 d.o.f. This is due to the strong variability in the light curve which was first found by \cite{HalpernGCN1} and is also discussed in \cite{Dai2007}. This fit, along with the residuals showing the strong variability, is shown in Fig. \ref{power-law}. The parameters we find for this fit ($\alpha_{\rm plateau}=0.288\pm0.026$, $\alpha_1=0.971\pm0.008$, $\alpha_2=2.524\pm0.052$, $t_{b1}=0.090\pm0.005$ days, $t_b=2.216\pm0.049$ days) are concurrent with those of \cite{Dai2007} (who find $\alpha_1\approx1.0$, $\alpha_2\approx2.9$ and $t_b\approx2.55$ days).

 Due to the high data density, we are able to let the break smoothness parameter $n$ vary for the second break (the first break had to be fixed to $n=10$), and our result ($n=4.3\pm0.7$) is in agreement with the tentative $\alpha_1-n$ correlation found by \cite{ZKK}. Still, the significant improvement shows that the light curve is basically a double smoothly-broken power law, and the steep late decay indicates that this break is a jet break, as first noted by \cite{ThoeneGCN} and also found by \cite{Dai2007}. The rest frame jet break time of 0.52 days is typical for the optical afterglows of the pre-\emph{Swift} GRB sample \citep{ZKK}.
 Jet breaks, a common feature in well-monitored pre-\emph{Swift} optical afterglows
\citep[e.g.,][]{ZKK}, have not been found in many \emph{Swift} afterglows\footnote{Note, though, that \cite{DaiDepp} argue that optical follow-up in the \emph{Swift} era, especially with UVOT only, often does not persist long enough (and to enough depth) to find jet breaks. }, especially in the X-rays
\citep[e.g.,][]{Mangano2007, Grupe2007, Sato2007, Racusin09}, and if there are breaks, then the
comparison between the optical and X-ray light curves show them to often be chromatic
\citep{Panaitescu2006, Oates2007}. Analysing both the optical and X-ray data of GRB 060526,
\cite{Dai2007} suggest that the (jet) break is achromatic.

\begin{figure*}
  \centering
 \includegraphics[width=17cm]{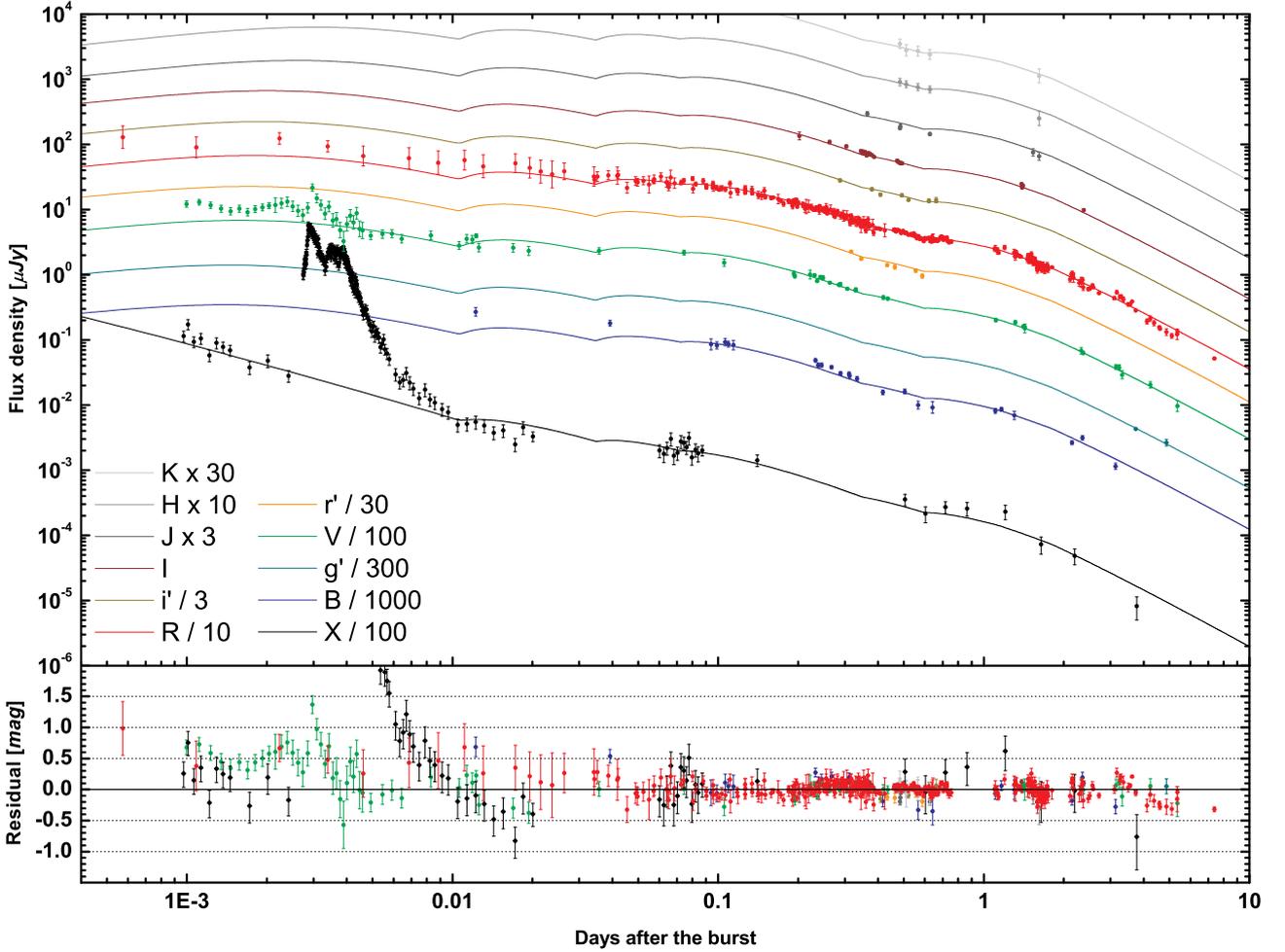}
   \caption{Fit to the light curve in X-ray and optical/NIR $Vr^\prime
	 Ri^\prime IJHK_S$ bands including a total of five energy injections.
	 Data before 500 s are excluded from the modelling, as well as the strong
	 X-ray flare. The light curves have been offset by constant factors as
	 given in the figure legend for better legibility. The $Bg^\prime$
	 bands are affected by additional Lyman forest absorption and were not
	 included in the fit. For the $Bg^\prime$ bands, the 
   model predictions have been multiplied by 0.50 and 0.55 respectively to match them with the data. The residuals are clearly improved in comparison to those plotted in Fig. \ref{power-law}, but short-timescale variations like the one at 0.7 days are still not fitted satisfactorily.}
              \label{Energy}
    \end{figure*}

If we take the X-ray data from
\cite{Dai2007} and fit it with the parameters of our double smoothly-broken power law fit using X-
ray data from 0.06 days onward, we concur that the fit is marginally acceptable, with $\csr1.6$
for 40 degrees of freedom. Using our own X-ray reduction, we obtain a worse result, with
$\csr1.7$ for 22 degrees of freedom. Similarly, with X-ray data from the \emph{Swift} XRT
repository \citep{Evans07, Evans09}, the fit is rejected ($\csr2.3$ for 33 degrees of freedom).
The reason is that there are less data points, but with smaller error bars, so the outliers are
weighted more strongly. A fit to the X-ray data alone results in a much earlier break time and
slopes that are less steep \citep[again in agreement with][]{Dai2007}, but we caution that the late
X-ray afterglow is only sparsely sampled and shows large scatter. For the three different reductions, we derive the following values using a smoothly-broken power law (a double-broken power 
law yields no statistical improvement), with $n=10$ fixed and no host galaxy:
$\alpha_1=0.92\pm0.05$, $\alpha_2=3.02\pm0.63$, $t_b=1.38\pm0.25$ (XRT repository light curve);
$\alpha_1=0.90\pm0.05$, $\alpha_2=2.78\pm0.61$, $t_b=1.34\pm0.28$ (our XRT reduction); $\alpha_1=-
0.29\pm0.56$, $\alpha_2=1.69\pm0.17$, $t_b=0.16\pm0.05$ (\citealt{Dai2007} data). As can be seen,
the latter fit is very different from the other two (which agree fully within error bars).
Even using data from 0.01 days onward (end of the X-ray flare), the fit is still significantly
different: $\alpha_1=0.43\pm0.06$, $\alpha_2=1.69\pm0.17$, $t_b=0.25\pm0.06$.

\subsection{Modelling the light curve with energy injections}
\label{EI}

Motivated by the similarities to such highly variable light curves as
that of the afterglow of GRB 021004, which was successfully modelled by
multiple energy injections (``refreshed shocks'') \citep{deUgarte2005},
we used the code of \cite{Johannesson06} to model the afterglow light curve. The code numerically solves the kinematic equations of an expanding shock front and calculates the resulting synchrotron emission. Relativistic effects are fully taken into account and the code supports delayed energy injection episodes.  Several energy injection 
episodes are applied as a possible 
scenario to explain the rebrightenings and shallow decay of the
afterglow. Preliminary results on a smaller data set were presented in
\cite{Johannesson09}. The number and a time range for the energy injections have to be inserted as initial guess for the fit and the fit then adjusts them to the best possible time within that range and determines the magnitude of the injection.
 As the very early data likely contain some
signature of the prompt emission, we exclude all data before 400 s as well as the very bright X-ray flare.  Due to Lyman forest
blanketing, data in $B$ and $g^\prime$ bands were excluded from the fit. The $V$
band is also affected by the blanketing, but was corrected with the
model from \cite{Madau1995} (brightened by 0.17 magnitudes) and included in the fit to have a better
spectral coverage at early times.

Figure \ref{Energy} shows the best
fit ($\csr2.0$) found using a model with a total of  five
energy injections: at 0.01, 0.04, 0.07, 0.35 and 0.60 days. Each energy injection episode adds 2 free parametres to the fit in addition to the 6 parametres needed for the standard afterglow model discussed below. The
initial energy injected into the outflow is\footnote{The error limits presented
here are $1\sigma$ estimates, but due to degeneracy in the model
parameters, the actual range of parameter values can be much larger.}
$E_0=5^{+2}_{-3}\times10^{49}$ ergs.
The energy injections then add 1.8, 4.1, 4.2, 8, and finally 14
times the initial energy release $E_0$ to the afterglow, for a total
energy release in the afterglow of $1.6^{+0.8}_{-1.1}\times10^{51}$ ergs.

The first three injections are responsible for the shallow
afterglow decay between 0.008 and 0.25 days.  The quality of the data
does not allow to discriminate between this three-injection
scenario and a continuous injection. Using a lower number of
injections does not yield a satisfactory fit to the observed data. Since
there are no direct indications of injections in the light curve, the
time of each of the three injections is not well determined.  A direct
consequence of this is that the energy of each individual injection in
this phase is not well determined, while the total energy released is
fairly consistent.  The time and energy of the last two injections are,
however, better constrained by the data.

Further results of the modelling are a high density of the circumburst
medium of $n_0=600^{+3000}_{-500}$ cm$^{-3}$ \citep[a high value,
comparable to the high-$z$ GRB 050904,][]{Frail2006}, a rather low
opening angle of $\theta_0 = 2^\circ\!\!.2^{+1^\circ\!\!.5}_{-1^\circ\!\!.8}$ and an
electron index of $p=2.1^{+0.2}_{-0.3}$ (host galaxy extinction is
assumed to be negligible, see \kref{SEDchap}).  The peak frequency
$\nu_m$ passes through the optical/NIR at very early times, while
the data are most likely affected by the prompt emission.  The cooling
break $\nu_c$ is between the optical and X-rays up to 6 days after the
burst, i.e., over the whole data span. Furthermore, we find
$\varepsilon_e=2^{+5}_{-1.5}\times10^{-2}$ and $\varepsilon_B=7^{+40}_{-6}\times10^{-5}$, with $\varepsilon_e$ being the fraction of the energy in the electron population and $\varepsilon_B$ the fraction of the energy in the magnetic field.  Note that the definition of $\varepsilon_e$ has been changed in the model from \cite{Johannesson06} to the definition of \cite{PK2001} to allow for $p<2$ in the model. This model reproduces the global properties of the optical/NIR light curves well, with the rather high $\chi^2$/d.o.f. $=2.0$ resulting from rapid variability in the late time afterglow. This is incompatible with the homogeneous shock front assumed in the numerical code.
The X-ray light curve is also fitted reasonably well.

\subsection{The spectral energy distribution and host extinction}
\label{SEDchap}

\begin{figure}
  \centering
 \includegraphics[width=\columnwidth]{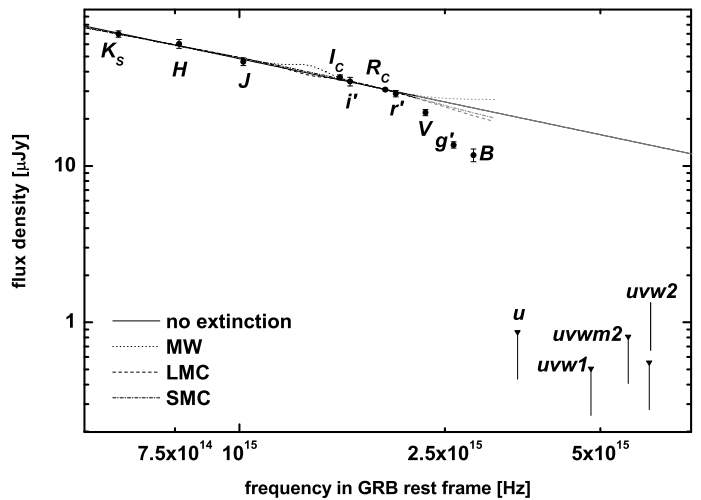}
   \caption{Spectral energy distribution of the afterglow of GRB 060526 in $uvw2\,uvm2\,uvw1\,uBg^\prime Vr^\prime R_Ci^\prime I_CJHK_S$, and fits with no extinction (straight black line), MW extinction (dotted line), LMC extinction (dashed line) and SMC extinction (thick dash-dotted line). The fit with MW dust finds (unphysical) negative extinction, causing an ``emissive component'' instead of a 2175 {\AA} absorption bump. Data beyond $2.2\times10^{15}$ Hz ($Vg^\prime B u\,uvw1\,uvm2\,uvw2$) were not included in the fit due to Lyman forest blanketing, the grey curves represent extrapolations. The extinction curves of \cite{Pei1992} are not correctly defined beyond $3.2\times10^{15}$ Hz. The UVOT UV filters are upper limits only, showing the strong flux decrease beyond the Lyman cutoff. The flux density scale is measured at the break time.}
              \label{SED}
    \end{figure}

\begin{table}
\caption{Fits to the spectral energy distribution of GRB 060526. Columns are the dust model, the goodness of the fit, the spectral slope and the derived extinction in the rest-frame $V$ band.}
\label{SEDtab}    
\centering                        
\begin{tabular}{l c c c}       
\hline\hline   
Dust & $\chi^2$/d.o.f. & $\beta$ & $A_V$ \\
\hline
none & 0.425	&	$0.695\pm0.035$ & ... \\
MW   & 0.309	&	$0.833\pm0.149$	&	$-0.126\pm0.132$ \\
LMC  & 0.466	&	$0.494\pm0.397$	&	$0.113\pm0.223$ \\
SMC  & 0.396	&	$0.552\pm0.198$	&	$0.055\pm0.075$ \\
\hline \hline
\end{tabular}
\end{table}

Following the procedures outlined in \cite{Kann2006}, we derive the optical spectral energy distribution (SED) of the GRB 060526 afterglow and fit it with several dust models \citep[Milky Way, MW; Large Magellanic Cloud, LMC; and Small Magellanic Cloud, SMC;][]{Pei1992} to derive the line-of-sight extinction in the host galaxy. Due to the strongly variable light curve, we choose the approach \cite{Kann2006} used for the SED of GRB 030329, and shift the other bands to the $R_C$-band zero point to derive the colours. With this method, we can also look for colour changes. There may be marginal variations in $B-R$, but this colour remains constant within conservative errors, and is more sensitive to Lyman forest blanketing, as different filters will suffer a different amount of blanketing.

The SED clearly shows the decreasing flux in the $B$, $g^\prime$ and $V$ bands, and especially in the $uvw2\;uvm2\;uvw1\;u$ bands, where only upper limits are found, due to the Lyman forest blanketing as well as the Lyman cutoff (see Fig. \ref{SED}), and we thus do not include these filters in our fit. Given the size of the errors, all fits, even with no extinction, are acceptable (see Table \ref{SEDtab}) and we are thus unable to prefer one dust model over another. The lack of $z$ band data does not allow us to constrain the existence of a 2175 {\AA} bump, therefore we have no evidence in favor of or ruling out MW and LMC dust. We do note, though, that MW dust leads to slightly negative extinction, while the LMC dust fit yields large errors. Thus, we henceforth use the SMC dust fit, as this is the most common dust type found in GRB host galaxies, both in pre-\emph{Swift} \citep{Kann2006, Starling07} and \emph{Swift}-era \citep{Schady07, Kann07} results. Note that for SMC dust, the extinction is 0 within errors as well.

Assuming the cooling break $\nu_c$ to lie blueward of the optical bands \citep[highly likely considering the X-ray spectral slope $\beta_X\approx1$,][ and the result of the numerical modelling]{Dai2007}, the standard fireball model gives for the electron power law index $p=2\beta+1=2.10\pm0.40$. This result (albeit with large errors) is in agreement with the result from the broadband modelling (\kref{EI}) and with the canonical $p=2.2$, and very similar to many other GRB afterglows \citep[e.g.][]{Kann2006, Starling07}.  
Our results are in contrast to \cite{Dai2007}, who derive a very steep slope $\beta_0=1.69^{+0.53}_{-0.49}$ from $Br^\prime i^\prime $ data only (correcting for Lyman absorption) and conclude that the optical and the X-ray data lie on the same slope.


\subsection{Host search}
\label{Host}

In the late-time VLT images, we do not detect any source at the position of the afterglow down to a limit of $R_C>27.1$ mag (which transforms into an absolute magnitude limit of M$_R>-20.1$). There is one source present at {$\sim$1\farcs5} South-East of the afterglow position with $R_C=26.4\pm0.2$, which at a redshift of $z=3.221$ would transform into a physical offset between GRB and host galaxy of $\sim11.5$ kpc. A long-slit spectrum covering the afterglow position and this galaxy does not show any trace at these positions. If this galaxy was associated with the GRB, it would be one of the largest offsets between a long-duration GRB and its host galaxy known \citep{Bloom2002}. The strong absorption lines from the ISM seen in the afterglow spectrum (see Sect. \ref{spectrum}) as well as the high circumburst density inferred from the numerical modeling (see Sect. \ref{EI}) do in fact not favour a large distance from the host galaxy. A position outside of their host galaxies had been inferred for GRB 070125 \citep{Cenko070125} and GRB 071003 \citep{Perley08b}, however, those spectra showed very weak absorption lines, contrary to what we observe for GRB 060526. Concluding, we have no direct spectroscopic evidence for or against an association of the GRB with the galaxy.

We also observed the field with ACS on {\it HST}. Astrometry was performed relative to our TNG image, yielding a position accurate to $\sim0.1$\arcsec. At the location of the burst we do not find any evidence for an underlying host galaxy.  To estimate the limits we place 50 apertures randomly on the sky and measure the standard deviation in their count rates. This implies that the underlying host galaxy of GRB 060526 is fainter than F775W(AB)$\,>\,$28.5 ($3 \sigma$) or an absolute magnitude of M(1500 \AA{}) $>$ --18.3\,mag which is fainter than 0.5 L* according to \cite{Gabash04}. The flux density at the host position is $0.004\pm0.005\;\mu$Jy.


Host galaxies of long-duration GRBs have often been found to be faint irregular galaxies \citep{Fruchter06, Christensen04} which are difficult to detect at higher redshifts. So far, there are only eight bursts with $z>3$ where the detection of a host galaxy has been published, namely GRB 971214 ($z=3.418$, \citealt{Kulkarni98}), GRB 000131 ($z=4.500$, \citealt{Andersen000131, Fruchter06}), GRB 030323 ($z=3.3718$, \citealt{Vreeswijk04}), GRB 060206 ($z=4.04795$, \citealt{Fynbo06a, Thoene060206, Chen09}), GRB 060210 ($z=3.9133$, \citealt{Fynbo09, Perley09}), GRB 060605 ($z=3.773$, \citealt{Ferrero08}), GRB 090205 ($z=4.6503$, \citealt{D'Avanzo090205}), and GRB 090323 ($z=3.568$, \citealt{CenkoLAT, McBreenLAT}). The host galaxies of GRB 020124 \citep{Berger2002, Chen09}, GRB 050730, GRB 050908, GRB 060607A, GRB 070721B \citep[all][]{Chen09}, GRB 050904 \citep{Berger2007} and GRB 060510B \citep{Perley09}, on the other hand, were not detected to very deep limits in the optical and the NIR. 

The distribution of pre-\emph{Swift} $R_C$ band host galaxy magnitudes peaks at $R_C(AB)=25$ \citep{Fruchter06} but extends out to 29 with a typical redshift $z\approx1.4$. \emph{Swift} GRBs (and thus their hosts) however have a higher mean redshift of $z=2.8$ \citep{Jakobsson06a}\footnote{An updated version of the redshift distribution can be found under http://raunvis.hi.is/$\sim$pja/GRBsample.html - as of July 1, 2010, the mean redshift has become lower, $z=2.19$, see also \cite{Fynbo09}, who find a mean and median of $z=2.2$.}, so the distribution will be shifted out to even fainter magnitudes. 
\cite{Ovaldsen07} also find a higher magnitude for \emph{Swift} hosts than for pre-\emph{Swift} bursts by comparing the expected detection rate from pre-\emph{Swift} hosts with detections and upper limits derived from imaging the fields of 24 \emph{Swift} and \emph{HETE II} bursts from 2005 -- 2006.


\section{Spectroscopy results}\label{spectrum}

\subsection{Line identification}

We detect a range of metal absorption lines as well as a Lyman limit system (LLS) originating in the host galaxy of GRB 060526. A redshift  of $z=3.221$ was determined in  \cite{Jakobsson06b} from a number of these absorption lines using the spectra taken with the 600V grism presented in this article. 

\begin{figure*}
  \centering
  \includegraphics[width=17cm]{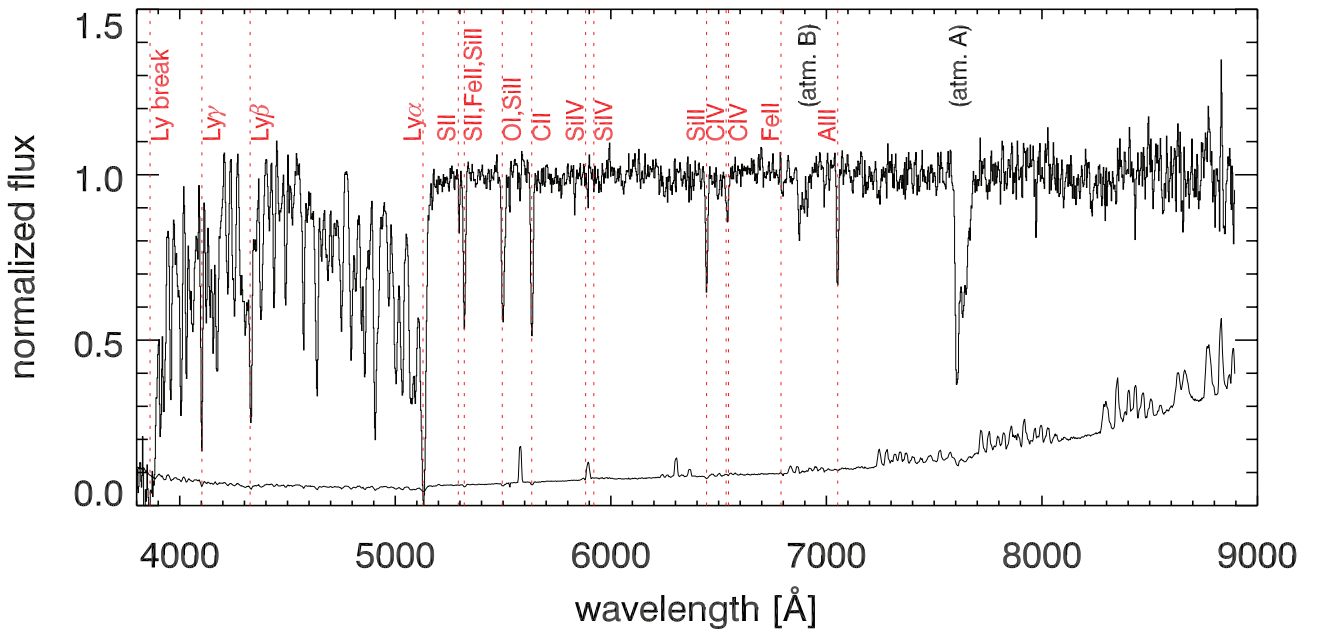}
   \caption{The spectrum taken with 300V grism, the two exposures have been combined to improve the S/N. The identified lines as well as the main atmospheric absorption bands are labelled. The line at the bottom shows the error spectrum.}
              \label{spec}
    \end{figure*}

Most of the lines were fitted from the combined 600V spectrum which covers all metal absorption lines detected longwards of Ly-$\alpha$, but provides a higher resolution than the 300V spectra. The 600I spectra only covers the AlII $\lambda$ 1670 line at the same resolution as the 600V grism. The range of 1200B is entirely within the Lyman-$\alpha$ forest but does not have a high enough resolution to deblend metal transitions from absorption caused by the Ly-$\alpha$ forest lines. We do, however, detect Ly-$\beta$ and Ly-$\gamma$ absorption in the 1200B grism. In the blue end of the 300V spectrum, one can clearly see the 915 \AA{} Lyman break at the redshift of the host galaxy. In Fig. \ref{spec}, we show the combined spectrum of the 300V grism with the identified lines indicated. We note that we do not detect any intervening system in the sightline towards GRB 060526, which is rather unusual for a GRB sightline, in particular at that redshift \citep[see, e.g.,][]{Prochter06}. Taking the strong \ion{Mg}{II} $\lambda$ 2796, 2803 doublet, the redshift path probed for any intervening system is between $z\sim0.8$ and 2.2. The $3\sigma$ limits on the non-detection of the \ion{Mg}{II} doublet vary between 0.17 {\AA} ($z=0.8$) and 0.82 {\AA} ($z=2.2$) (rest frame).

In order to determine the equivalent width (EW) of the strong absorption lines, we fitted the continuum around the lines in regions that were free of absorption and summed over the absorption contained within two times the full-width at half-maximum (FWHM) of the lines. For weak lines, we obtained better results due to the low S/N by fitting Gaussians. For this fit, we used a modified version of the gaussfit procedure provided in IDL\footnote{Available at http://www.pa.iasf.cnr.it/$\sim$nicastro/IDL/Lib/gfit.pro} which is more reliable in determining the continuum and fitting the actual line even if it is slightly blended with a neighbouring line. The upper limits on the EWs for a range of ions noted in Table \ref{lines} was determined from the spectra taken with the 300V grism due to the better S/N of those spectra. Between the individual spectra, we do not find any variability in the EW of the individual absorption lines.

\ion{S}{II} $\lambda$ 1259, \ion{Si}{II} $\lambda$ 1260 and \ion{Fe}{II} $\lambda$ 1260 are blended and cannot be fitted separately. We therefore cannot consider them for the derivation of the column density from the curve of growth fit as described below and only give the total EW in Table \ref{lines}. In contrast to what is noted in \cite{Jakobsson06b}, we cannot reliably detect any fine-structure lines and only give an upper limit for \ion{Si}{II*}. Fine-structure lines would be a clear indication that the detected absorption lines indeed originate in the host galaxy of the GRB, as they are assumed to be produced by UV pumping from the afterglow \citep[e.g.][]{Prochaska051111, Vreeswijk07, Delia09}. The redshift derived is therefore to be strictly taken as a lower limit only, the detection of the Lyman $\alpha$ forest redward of the proposed redshift, however, excludes a significantly higher redshift for the burst.

\begin{table*}           
\caption{EWs of detected absorption lines and 2$\sigma$ upper limits on some undetected lines, the EW for the blended systems include the contributions from all lines. The column densities were derived from CoG fitted for \ion{S}{II}, \ion{Si}{II}, \ion{Fe}{II}, \ion{C}{IV}, \ion{Si}{IV} and \ion{Al}{II}. \ion{Si}{II*} denotes a fine-structure line. Upper and lower limits were determined by assuming the ions to lie on the linear part of the CoG. The Ly$\alpha$ column density is taken from \cite{Jakobsson06b} and is based on the 600V grism. \label{lines}   }
\centering                        
\begin{tabular}{l l l l l l}       
\hline\hline
$\lambda_{\rm obs}$ & $\lambda_{\rm rest}$&ID& $z$ & EW$_\mathrm{rest}$& log N/cm$^{-2}$\\
 $[$\AA$]$ & $[$\AA$]$ & &  &$[$\AA$]$ & \\
\hline
4102.26	&972.54	&Ly$\gamma$ & 3.218& $2.39\pm0.10$			& ---\\
4329.70	&1025.72	&Ly$\beta$ & 3.221	& $2.92\pm0.06$			& ---\\
5131  	&1215.67	&Ly$\alpha$& 3.221	& ---						&$20.00\pm0.15$\\
(5254)	&1242.80	&\ion{N}{V}	&(3.221)	& $<0.09$				&$<13.66$\\
(5276)	&1250.58	&\ion{S}{II}	&(3.221)	& $<0.08$				&$<14.71$\\
5291.01	&1253.81	&\ion{S}{II}	&3.220	& $0.05\pm0.02$		&$14.58\pm0.25$\\
5320.00  	&1259.52	&\ion{S}{II}	&3.221	& ($1.32\pm0.07$)		& ---\\
		&1260.42	&\ion{Si}{II}	&3.221	& {\it blended}			& ---\\
		&1260.53	&\ion{Fe}{II}	&3.221	&					& ---\\
(5335)	&1264.74	&\ion{Si}{II*}	&(3.221)	& $<0.10$				&$<12.76$\\	
5497.30  	&1302.17	&\ion{O}{I}		&3.222	& $1.07\pm0.05$		&$>15.15$\\
5506.82	&1304.37	&\ion{Si}{II}	&3.222	& $0.70\pm0.21$		&$15.87\pm0.16$\\
(5559)	&1317.22	&\ion{Ni}{II}	&(3.221)	& $<0.11$				&$<13.59$\\
5634.24 	&1334.53	&\ion{C}{II}	&3.222	& $1.51\pm0.10$	 	&$>15.55$\\
5884.87  	&1393.76 	&\ion{Si}{IV}	&3.222	& $0.20\pm0.08$		&$13.50\pm0.14$\\
5923.05  	&1402.77	&\ion{Si}{IV}	&3.222	& $0.12\pm0.05$		&$13.50\pm0.14$\\
6444.50  	&1526.71 &\ion{Si}{II}	&3.221	& $0.86\pm0.04$		&$15.87\pm0.16$\\
6539.58   	&1548.20	&\ion{C}{IV}	&3.223	& $0.26\pm0.08$		&$14.10\pm0.13$\\
6546.39  	&1550.78 	&\ion{C}{IV}	&3.221	& $0.22\pm0.09$		&$14.10\pm0.13$\\
6790.24   	&1608.45 &\ion{Fe}{II}	&3.221	& $0.17\pm0.03$		&$14.28\pm0.24$\\
6800.38	&1611.20	&\ion{Fe}{II}	&3.221	& $<0.01$				&---\\
7052.62  	&1670.79	&\ion{Al}{II}	&3.221	& $1.02\pm0.05$		&$15.12\pm0.18$\\
(7863)	&1862.79	&\ion{Al}{III}	&(3.221)	& $<0.35$				& $<12.71$\\
(8552)	&2026.13	&\ion{Zn}{II}	&(3.221)	& $<0.45$				& $<12.73$\\
(8679)	&2056.25	&\ion{Cr}{II}	&(3.221)	& $<0.54$				& $<13.47$\\
\hline \hline
\end{tabular}
\end{table*}

\subsection{Column densities from curve of growth analysis}
Some of the strong absorption lines are saturated, which is a problem in low resolution spectra as the damping wings are not resolved and Voigt profile (VP) fitting cannot be adopted to derive a reliable column density. Furthermore, high resolution spectra of GRBs \citep{ProchaskaCoG} have shown that the strong metal absorption lines unresolved in low resolution spectra usually consist of a number of narrow, unsaturated components that would allow an accurate determination of the column density by fitting the different components separately.

\begin{figure}
  \centering
  \includegraphics[width=\columnwidth]{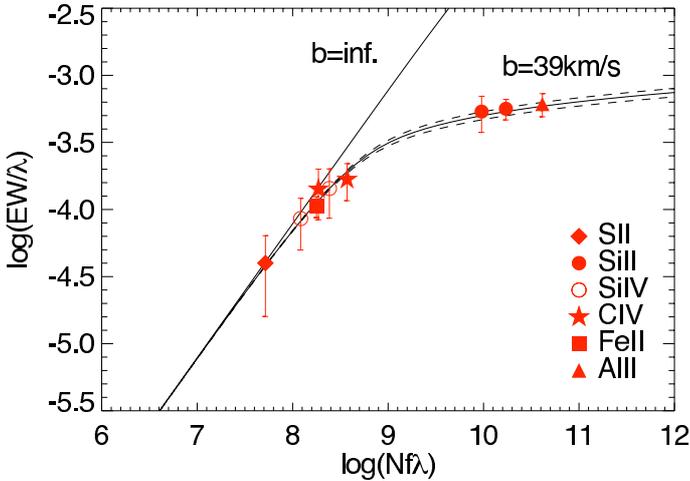}
   \caption{Multi-ion single-component curve of growth fit for the absorption systems in GRB 060526 using 6 different ions. The CoG is expressed in units of the column density ``N'', the oscillator strength of the transition ``f'' and the wavelength of the transition. As comparison, $b = \infty$ is plotted, the dotted line marks the 1 $\sigma$ deviation from the fit.}
              \label{CoG}
    \end{figure}

If only low resolution spectra are available, one has to adopt a curve of growth (CoG) analysis \citep{Spitzer78, SavageSembach96} which directly relates the EW to the column density on the linear part of the CoG where the lines are not saturated (optical depth $\tau_0<1$), but depends on the Doppler $b$ parameter of the medium on the flat part of the curve which applies for mildly saturated lines. For GRBs, one usually has to do a multiple-ion single-component CoG (MISC-CoG) analysis, adopting the same effective Doppler parameter for all ions. Here, we used all unblended ionic lines that were not heavily saturated, namely \ion{S}{II}, \ion{Si}{II}, \ion{Fe}{II}, \ion{Al}{II} as well as \ion{C}{IV} and \ion{Si}{IV} and calculate the $\chi^2$ minimum going through the parameter space for the column densities of each ion and a range of $b$ parameters. We then find the best fit for $b=39\pm3$ km s$^{-1}$ (see Fig. \ref{CoG}) and the column densities as noted in Table \ref{lines}. Most of the ions lie near the linear part of the CoG and are therefore rather independent of $b$, \ion{Si}{II} and \ion{Al}{II} lie on the flat part of the CoG and the column densities are only noted as lower limits as derived from the linear part of the CoG. We excluded the saturated \ion{O}{I} and \ion{C}{II} transitions from the fit and only list lower limits for these two ions for which we take the column density resulting from the linear part of the CoG. We also discarded the blended absorption lines of \ion{Si}{II} $\lambda$ 1260, \ion{S}{II} and \ion{Fe}{II} even though \ion{Si}{II} is the dominant contribution to the absorption since we detect two other unblended \ion{Si}{II} transitions at $\lambda$ 1304 and $\lambda$ 1526 \AA{}.

There are several problems connected with the use of the CoG that have to be considered. Different ionisation levels should actually be treated in separate analyses as they might occur in different regions in the absorbing system. However, we assume that the absorption takes place in a relatively small region of the host galaxy and the resolution of the spectrum does resolve different components of the absorbing material in the host galaxy. Therefore, we also fit the higher ionisation levels of \ion{Si}{IV} and \ion{C}{IV} in the same CoG. Furthermore, they lie close to the linear part of the CoG and excluding them from the common fit would not change the derived $b$ parameter very much. Another problem with doing multiple-ion CoG analysis using strong lines in low-resolution spectra has been noted by \cite{ProchaskaCoG}, who compared column densities derived from low resolution spectra and CoG with high resolution data and VP fitting from the same bursts. He found that when including saturated lines in the fit, column densities are generally underestimated. Indications for that are if an effective $b$ parameter of $\gg$ 20 km s$^{-1}$ is found, since strong lines actually consist of a range of components with $b<20$ km s$^{-1}$. \cite{Savaglio06} however performed a similar analysis using CoG and the apparent optical depth (AOD) method as described in \cite{Pettini02}, which can be applied for medium resolution high S/N spectra, and found a good agreement between the two methods. The column densities of the saturated lines \ion{Al}{II} and \ion{Si}{II} are infact rather sensitive to the adopted $b$ parameter. These lines very likely consist of a number of unresolved weaker components which would lie on the linear part of the CoG, the real errors should therefore be larger but are difficult to estimate. 

\subsection{Metallicity and relative abundances}
Absorption lines that are likely not affected by dust depletion can be used to derive a metallicity of the medium in the line of sight to the GRB. The least dust-depleted element is Zn, which is, however, undetected in our spectra. We then use the relatively weak lines \ion{S}{II} and \ion{Fe}{II} derive relative metallicities compared to the hydrogen density with [M/H]=log(N$_M$/N$_H$)--log(N$_{M}$/N$_{H}$)$_\odot$ using solar abundances from \cite{Asplund05}. Here we derive metallicities of $[{\rm S}/{\rm H}]=-0.57\pm0.25$ and $[{\rm Fe}/{\rm H}]=-1.22\pm0.24$.

Fe is usually affected by dust depletion \citep{SavageSembach96} and corrections have to be adopted. Using the relation between the Zn and Fe abundance in \cite{Savaglio06}, we find a metallicity of $[{\rm Fe}/{\rm H}]=-1.09\pm0.24$ which marginally agrees within the errors with the value derived from sulphur. The only detected and unblended \ion{S}{II} line at $\lambda$ 1253, however, is only marginally detected and therefore the EW has large errors. The \ion{Fe}{II} doublet taken for the CoG fit, in contrast, is also slightly blended, but the fit of the stronger component can be considered as reliable. Despite the dust depletion, the metallicity derived from Fe might be the most reliable one in this case and we therefore assume a metallicity of [Fe/H] = --\,1.09 for the host of GRB\,060526.  

Independent of the ion used, the metallicity is rather high compared to other galaxies at redshift z $\approx$ 3, but among the typical metallicities derived for other GRB hosts \citep{Fynbo06a} at that redshift. For those measurements, different ions have been used depending on the quality and the wavelength coverage of the spectra. Our results show that caution is required when comparing the metallicities derived from different elements, as they might be differently affected by dust depletion and/or evolution. This is especially true when saturated lines in low resolution spectra are used to derive the column densities and hence the abundances as it is the case for, e.g., GRB 000926 \citep{Savaglio03} and GRB 011211 \citep{Vreeswijk06}. For GRB\,050401 \citep{Watson04} and GRB 050505 \citep{Berger05}, the authors themselves note that due to saturation the reported metallicity is indeed a lower limit. 

Our spectra do not allow us to determine a dust depletion pattern from the relative abundances of heavier elements, since at least four elements out of Zn, Si, Mn, Cr, Fe, Ni or S are necessary to do such a fit. The difference in the relative abundances between S, Si and Fe might suggest the presence of some dust in the line-of-sight towards the GRB, however, large extinction is excluded from the afterglow SED. On the other hand, this difference could also be due to an enhancement in $\alpha$ element production\footnote{$\alpha$ elements are produced in massive, metal-poor stars through the $\alpha$ process and include elements with integer multiples of the He nucleus mass such as O, Si, S, Ca, Mg and Ti.} which is likely to happen in the young star-forming host of a GRB. Generally, GRB hosts have higher $\alpha$/Fe ratios than QSO-DLAs \citep{Prochaska07}, which can either be interpreted as large dust depletion consistent with the higher metallicities of GRB sightlines or as $\alpha$ element enhancement. There are indications that most of the $\alpha$/Fe ratio is due to dust depletion traced by a large [Zn/Fe] or $[{\rm Ti}/{\rm Fe}]<0$. The latter is assumed to be a clear indicator of dust depletion ($[{\rm Ti}/{\rm Fe}]<0$) vs. $\alpha$ enhancement ($[{\rm Ti}/{\rm Fe}]>0$) \citep{Dessauges02}. Both elements are, however, not detected and the limit on Zn does not lead to a strong constraint on the [Zn/Fe] ratio to settle this issue in the case of GRB 060526. 

The ratio between high and low ionisation species in the spectrum clearly shows that most of the material in the line of sight is in a low ionization state. \ion{Si}{IV} and \ion{C}{IV} have rather low column densities and lie close to the linear part of the CoG in contrast to their low ionization species \ion{Si}{II} and \ion{C}{II} which are both saturated, whereas \ion{Al}{III} is not even detected in our spectra. We then derive column density ratios of log(\ion{Si}{IV}/\ion{Si}{II}) $=-2.37$, log(\ion{C}{IV}/\ion{C}{II}) $<-1.45$ and log(\ion{Al}{III}/\ion{Al}{II}) $<-2.41$. From the large sample of \emph{Swift} long GRB afterglow spectra \citep{Fynbo09}, this seems fairly normal for an average GRB sightline. However, the few GRBs occurring in LLSs ($\log N_{\rm HI}/{\rm cm}^{-2} < 20.3$) usually show a higher fraction of ionized material compared to GRB-DLA sightlines. This might either be due to decreased shielding of the highly ionizing afterglow flux by the lower hydrogen column density, by a rather special arrangement of the GRB inside the host galaxy or simply by a small host galaxy. Since our spectra show a low fraction of highly ionized material this implies that the absorbing material has a relatively large distance from the GRB.

We do not detect \ion{N}{V} in our spectra but provide an upper limit of $\log N < 13.66$. \ion{N}{V} was detected in only four sightlines towards GRBs \citep{Prochaska08} and likely traces the immediate environment of the GRB as it has a high ionization potential and requires a strong radiation field. Our upper limit is lower than the column densities for those GRB sightlines where \ion{N}{V} could be detected which again implies that the absorbing gas probed by our spectra are most likely not close to the GRB itself. Furthermore, we do not detect any fine structure lines and only derive an upper limit on \ion{Si}{II*} of $\log N/{\rm cm}^{-2} < 12.76$. Fine structure lines are assumed to be pumped by the UV radiation field of the GRB \citep{Vreeswijk07} which also indicates that the gas is likely very far from the GRB itself. \ion{Si}{II*} requires a less strong radiation field and has been detected in Lyman break galaxies \citep{Pettini02} where the UV radiation from young stars provides the necessary radiation field.

\section{Discussion and conclusions}
GRB 060526 had a relatively bright afterglow that allowed us to obtain a solid dataset, both photometrically and spectroscopically. We achieved a dense light curve coverage over several days which allowed a detailed study of the afterglow properties, and obtained a series of low resolution but high signal-to-noise spectra to study the host environment. 

The optical light curve can be fitted with a double smoothly-broken power law with a breaks at $t_{b1}=0.090\pm0.005$ and $t_b=2.216\pm0.049$ days, and decay slopes of $\alpha_{plateau}=0.288\pm0.026$, $\alpha_1=0.971\pm0.008$, and $\alpha_2=2.524\pm0.052$. The dense sampling of especially the $R_C$-band light curve also reveals additional variability on top of the power laws. These features could be explained either by extended activity of the central engine or through interactions of the shock with the interstellar medium. For the case that the variability arises from external shocks, several mechanisms have been considered. In GRB 021004, both density variations of the external medium into which the GRB jet plows and angular inhomogenities of the jet surface were considered \citep{NPG03}. However, \cite{Nakar07} show that density variations would cause much smaller fluctuations than those observed in GRB afterglows and can therefore be ruled out. Another possibility is the injection of additional energy into the shock by slower shells that catch up with the shocked region as it decelerates, this model was used successfully to describe GRB 021004 \citep{deUgarte2005} and also works better than two other models (double jet and density fluctuations) to describe the highly complex light curve of GRB 030329 \citep{Huang2006}. Thus, variability can give either information on the medium surrounding the GRB or on the activity of the central engine. A more intriguing possibility is that the flares may be emitted from another region closer to the central engine, resulting from late internal shocks. Powerful X-ray flares that are attributed to late central engine activity have been observed in about 50\% of all \emph{Swift} GRBs \citep[e.g.][]{Burrows2005, Chincarini2007, Krimm2007}, and strong optical/NIR flaring contemporaneous with the GRB prompt emission may also occur \citep{Vestrand041219A, Blake041219A, Vestrand050820A, Racusin08}, thus making optical flares from late central engine activity an interesting prospect \citep{KannGCN, Malesani2007}.

Indeed, \cite{Dai2007} have suggested that the optical variability of the afterglow of GRB 060526 is due to flares from late internal shocks (the very early rapid optical variability we present here is very probably due to central engine activity, as it is seen contemporaneously in gamma and X-rays). \cite{Khamitov08}, on the other hand, conclude that the short timescale of the variabilities requires the jet to be non-relativistic already at $\sim1$ day and could then be explained by external density fluctuations. Our analysis lends tentative support to the notion of flares from internal shocks, finding decay slopes for two flares that exceed what should be possible from external shocks. But we caution that the errors of these fits are large due to a low amount of data in the decaying parts. Furthermore, globally, a model using refreshed external shocks is able to account for the light curve variations, although microvariability remains. This creates the intriguing possibility of reverberation effects \citep[see][for a case of reverberation between gamma-rays and optical emission]{Vestrand050820A}. Short flares in the X-ray or optical bands signal internal shocks from long-term central engine activity, and when these shells catch up with the forward shock front, they re-energise the external forward shock. The detection of such behaviour would probably require dense multi-band observations of a bright afterglow to search for SED changes at high time resolution combined with detailed modelling of the data. This way, one could discern between internal shocks (which are expected to have a different spectral index from the forward shock afterglow) and refreshed, external shocks (which are achromatic). Our data set of the afterglow of GRB 060526 does not allow us such a detailed decomposition.

From the analysis of our low resolution spectra with different resolutions, we detect a LLS and a number of metal absorption lines that all lie at a redshift of $z=3.221$. The low resolution only allows us to derive column densities from measuring the EWs of the absorption lines and adopt a MISC-CoG analysis where we exclude the most saturated as well as blended transitions. We find a best fit for the Doppler parameter of $b=39\pm3$ km s$^{-1}$ and most of the ions used for the fit lie on the linear part of the CoG which allows a relatively reliable determination of the column densities. The relative abundances of different metals in the spectra indicate some dust extinction, but an intrinsic difference due to enhancement of the production of certain elements cannot be excluded. The very low amount of dust detected in the afterglow SED may indicate that the latter might be the favored possibility. 

The column density of neutral hydrogen is rather low compared to other GRBs. We derive a metallicity for the host of $[{\rm Fe}/{\rm H}]=-1.09$ which is slightly higher than metallicities determined from other GRB afterglow spectra. According to the definition of QSO absorbers, the host of GRB 060526 is classified as a LLS ($19 < \log N_{\rm HI}/{\rm cm}^{-2} < 20.3$), which seem to have on average higher metallicities than damped Lyman $\alpha$ systems (DLA; \citealt{Peroux07}) and a steeper evolution towards lower redshifts. Around redshift 3, however, the metallicities of both samples are within the same range. Also, GRB hosts show a trend towards increasing metallicity with lower redshifts \citep{Fynbo06a, Savaglio06}. Taking into account that most of the sample used only low-resolution spectra to derive the metallicity (which only gives lower limits for the column densities and the metallicity) this evolution might, however, not be as pronounced as for DLAs and LLS. This might imply that the enrichment of the ISM in the early universe had taken place at earlier times than assumed. The absorbing material along the line-of-sight is mostly in the neutral state, as usually observed for long GRB-DLAs, while sightlines with lower $\log N_{\rm HI}$ often contain more ionized material. This might either imply that we have a very small host galaxy or that the GRB is placed somewhere in the outskirts of its host. In general, the low ionization points to a relatively large distance of the absorbing material from the GRB itself.

There is no underlying host galaxy of GRB\,060526 detected down to a deep limit of 28.5 mag (in F775W AB) in HST/ACS data. At that redshift, this means the host has an absolute magnitude M(1500\AA{}) $>-18.3$ mag, fainter than an 0.5 L* galaxy at that redshift. Long GRBs have been found to occur in actively star-forming galaxies and star formation is assumed to shift towards smaller and fainter galaxies over time \citep[e.g.][]{Cowie96} while massive galaxies prove to be rather unchanged throughout the history of the universe \citep[e.g.][]{Abraham99, Heavens04}. One would therefore expect that GRB hosts should also have higher luminosities towards higher redshifts. \cite{Fynbo08} concluded that the observed metallicity distribution of GRB hosts (as well as QSO absorbers) at $z\approx3$ can be explained by the luminosity function of galaxies at that redshift and assuming a luminosity-metallicity relation as derived for other high-redshift samples \citep{Ledoux06, Erb06}. The non-detection of the host of GRB\,060526 down to deep limits, however, would not support this suggested evolution. The data neither strongly support nor allow us to rule out that the GRB is associated with the nearby galaxy. While the offset would be very large compared to typical long GRB offsets, it is possible the burst occurred in a locally dense star-forming region which is not detected even in our very deep imaging.

\acknowledgements
We thank the referees for constructive comments that improved the paper. CCT wants to thank C\'edric Ledoux for comments on the curve-of-growth analysis and S. Campana for comments on the text. We thank the support astronomer and staff at the Nordic Optical Telescope for obtaining the observations.\\
The Dark Cosmology Centre is funded by the DNRF. HD acknowledges support from the Research Council of Norway. LH acknowledges support from Science Foundation Ireland grant 07/RFP/PHYF295. PJ acknowledges support by a Marie Curie European Reintegration Grant within the 7th European Community Framework Program and a Grant of Excellence from the Icelandic Research Fund.  IFB acknowledges partial support by the RFFI grant ''09-02-97013-povolzhe''. PAC gratefully acknowledges the support of NWO under grant 639.043.302. CB, AG, and GG acknowledge the University of Bologna for the funds Progetti di Ricerca Pluriennali. SF acknowledges the support of the Irish Research Council for Science, Engineering and Technology, cofunded by Marie Curie Actions under FP7.\\
Based on observations made with the Nordic Optical Telescope, operated on the island of La Palma jointly by Denmark, Finland, Iceland, Norway, and Sweden, with the William Herschel Telescope and the Telescopio Nazionale Galileo in the Spanish Observatorio del Roque de los Muchachos of the Instituto de Astrof\'isica de Canarias and at the Astronomical Observatory of Bologna in Loiano (Italy). Collection of SMARTS data is supported by NSF-AST 0707627. MIRO is supported by the Department of Space, Govt. of India. The Peters Automated Infrared Imaging Telescope (PAIRITEL) is operated by the Smithsonian Astrophysical Observatory (SAO) and was made possible by a grant from the Harvard University Milton Fund, the camera loan from the University of Virginia, and the continued support of the SAO and UC Berkeley. The PAIRITEL project and JSB are further supported by NASA/Swift Guest Investigator Grant NNG06GH50G. We thank M. Skrutskie for his continued support of the PAIRITEL project. The W.M. Keck Observatory is operated as a scientific partnership among the California Institute of Technology, the University of California and the National Aeronautics and Space Administration and  was made possible by the generous financial support of the W.M. Keck Foundation. IFB, RAB, IMKh, MNP, RAS extend thanks to T\"UB\.ITAK, IKI, and KSU for a partial supports in using the RTT150 (Russian-Turkish 1.5-m telescope in Antalya).


\end{document}